\documentclass[fleqn,usenatbib]{mnras}
\usepackage{newtxtext,newtxmath}
\usepackage{caption}

\usepackage[T1]{fontenc}

\DeclareRobustCommand{\VAN}[3]{#2}
\let\VANthebibliography\thebibliography
\def\thebibliography{\DeclareRobustCommand{\VAN}[3]{##3}\VANthebibliography}
\usepackage{rotating}

\usepackage{graphicx}	
\usepackage{amsmath}	
\usepackage{lipsum}  
\usepackage{subfigure}

\title[X-ray Simulations of Polar Gas]{X-ray Simulations of Polar Gas in Accreting Supermassive Black Holes}

\author[McKaig et al.]{
Jeffrey McKaig$^{1}$\thanks{E-mail: jmckaig@gmu.edu},
Claudio Ricci$^{1,2,3}$,
St\'ephane Paltani$^{4}$,
Shobita Satyapal$^{1}$
\\
$^{1}$George Mason University, Department of Physics and Astronomy, MS3F3, 4400 University Drive, Fairfax, VA 22030, USA\\
$^{2}$N\'ucleo de Astronom\'ia de la Facultad de Ingenier\'ia Universidad Diego Portales, Av. Ej\'ercito Libertador 441, Santiago, Chile\\
$^{3}$Kavli Institute for Astronomy and Astrophysics, Peking University, Beijing 100871, China\\
$^{4}$Department of Astronomy, University of Geneva, 1290 Versoix, Switzerland
}

\date{Accepted 2021 October 26. Received 2021 October 24; in original form 2021 August 24}

\pubyear{2020}

\begin{document}
\label{firstpage}
\pagerange{\pageref{firstpage}--\pageref{lastpage}}
\maketitle

\begin{abstract}
\indent Recent observations have shown that a large portion of the mid--infrared (MIR) spectrum of active galactic nuclei (AGN) stems from the polar regions. In this paper, we investigate the effects of this polar gas on the X-ray spectrum of AGN using ray-tracing simulations. Two geometries for the polar gas are considered, (1) a hollow cone corresponding to the best fit MIR model and (2) a filled cone, both with varying column densities (ranging from $10^{21}-10^{22.5}$\,cm$^{-2}$) along with a torus surrounding the central X-ray source. We find that the polar gas leads to an increase in the equivalent width of several fluorescence lines below $\sim 5$\,keV (e.g., O, Ne, Mg, Si). A filled geometry is unlikely for the polar component, as the X-ray spectra of many Type 1 AGN would show signatures of obscuration. We also consider extra emission from the narrow line region such as a scattered power-law with many photoionised lines from obscured AGNs, and different opening angles and matter compositions for the hollow cone. These simulations will provide a fundamental benchmark for current and future high spectral resolution X-ray instruments, such as those on board {\it XRISM} and {\it Athena}. 

\end{abstract}




\section{Introduction}

    Supermassive black holes (SMBHs) are thought to inhabit the center of every massive galaxy (e.g., \citealp{1998AJ....115.2285M}). Throughout its lifetime, a SMBH can accrete significant amounts of matter, creating a very luminous and compact source of photons through the formation of an accretion disk \citep{1973A&A....24..337S}. When a SMBH is undergoing this accreting phase, it is classified as an active galactic nucleus (AGN). AGN can then be subdivided based on their optical, radio, and X-ray properties. In order to explain the differences in the observational properties of accreting SMBHs, particularly in the optical and X-ray band, the unified model of AGN proposes a large scale dusty torus which sourrounds the broad line region and the accreting SMBH \citep{1993ARA&A..31..473A,1995PASP..107..803U, 2015ARA&A..53..365N, 2017NatAs...1..679R}. The differences in AGN are then explained by the observers viewing angle with respect to the torus. When viewing an AGN through the torus, the central SMBH and accretion disk are obscured, and narrow emission lines are present. Observing an AGN perpendicular to the plane of the torus leads to an unobscured line of sight of the inner regions of the accreting system, which allows for the detection of both broad and narrow emission lines. 

    In AGN, X-rays are produced in a small region close to the SMBH (e.g., \citealp{2009ApJ...693..174C, 2014A&ARv..22...72U}). The accretion disk cannot be responsible for the observed X-ray radiation because the temperature of the disk is not sufficient to create X-ray photons \citep{1973A&A....24..337S}. Instead, it is now widely accepted that optical/UV photons from the disk are up-scattered through the process of inverse Compton scattering when interacting with heated electrons in the X--ray corona \citep{1991ApJ...380L..51H, 1993ApJ...413..507H}. The process in which these electrons become heated is still debated and could be caused by magnetic flares (e.g., \citealp{1994ApJ...432L..95H,2000MNRAS.318L..15M}), clumpy disks (e.g., \citealp{1988MNRAS.233..475G}), or aborted jets \citep{1997A&A...326...87H}. The primary component of the X-ray continuum observed in AGN can be approximated by a power-law with an exponential cutoff at $\sim 200$ keV (e.g., \citealp{10.1093/mnras/sty1879}). The slope of this power-law is determined by the photon index $\Gamma$, which typically has a value between 1.8 and 1.9 \citep{Nandra7469,10.1111/j.1365-2966.2011.18207.x, 2017ApJS..233...17R}. Absorption in the X-rays is driven by photoelectric absorption and Compton scattering, the latter becoming important at column densities of $~1/\sigma_{\mathrm{T}}\approx 10^{24}$ cm$^{-2}$, where $\sigma_{\text{T}}$ is the Thomson cross section. AGN that are observed through column densities higher than this threshold are dubbed Compton thick (CT; e.g., \citealp{10.1046/j.1365-8711.2000.03721.x,10.1111/j.1365-2966.2010.17902.x,10.1111/j.1365-2966.2012.20908.x,2015ApJ...815L..13R, 10.1093/mnras/stw1764, Hikitani_2018}). Reprocessed radiation is responsible for many characteristic features in the X-ray spectra of AGN, such as the reflection hump at $\sim 30$\,keV and the Iron K$\alpha$ line at 6.4\,keV. 
    
    Until recently, most of the mid-infrared (MIR) emission associated to AGN has been thought to have its origin in the torus \citep{2008ApJ...685..147N,2012MNRAS.420.2756S}. \citet{Jaffe2004} reported MIR observations of NGC-1068 showing a warm dusty structure on scales of parsecs which surrounded a smaller hot structure. This was a significant confirmation of the presence of a torus-like structure in an AGN. Now, recent observations from the MID-infrared Interferometric instrument (MIDI) on the Very Large Telescope (VLT) show a large portion of the MIR spectrum of AGN stemming from the polar regions on scales of tens to hundreds of parsecs (e.g., \citealp{H_nig_2012}) and not only from the torus as previously thought (\citealp{10.1093/mnras/stx2227, 10.1093/mnras/stz220} and references therein). These observations have been made both by MIR interferometric (e.g., \citealp{2012ApJ...755..149H, 2013A&A...558A.149B, refId03, Leftley_2019}) and single dish observations (e.g., \citealp{Asmus_2016}). In addition to this, ALMA is also revealing polar outflows on the same spatial scales (e.g., \citealp{2016ApJ...829L...7G}) which suggests the torus may be an obscuring outflow. This is in contrast with the traditional unified model, which does not predict this elongated polar MIR emission. This polar gas (or component) has a significant effect on the MIR SED, but it is likely optically thin, otherwise the X-ray emission would be obscured for type 1 AGN \citep{Liu2019XraySO}. The origin of this polar component is still unknown. A plausible explanation is that this polar gas is the result of radiation pressure on dust grains from strong UV emission in the polar region from the accretion disk causing a dusty wind \citep{Ricci2017, Leftley_2019, 2019ApJ...884..171H, 2020ApJ...900..174V}, a scenario also consistent with hydrodynamical simulations (e.g., \citealp{2012ApJ...758...66W}).   

    The two nearest sources in which this polar component has been confirmed by interferometric studies are NGC-1068 \citep{refId01} and the Circinus galaxy \citep{refId02}, although such elongated emission has also been observed in several other sources (e.g., \citealp{refId03, 10.1117/12.2231077}). There has been extensive literature studying the effect of this polar gas on the MIR spectrum, as well as the effect of different distributions of the polar gas itself (e.g., \citealp{10.1093/mnras/stx2227, 10.1093/mnras/stz220, 10.1093/mnras/stz2289}). These studies concluded that the best model for the distribution of this polar gas to fit the MIR spectrum is a hollow cone or a hyperbolic cone on large scales, accompanied by a disk component on smaller scales. The hollow nature of these geometries is naturally consistent with the absence of obscuring material along the line of sight to Type 1 AGNs (e.g., \citealp{Ricci2017, 2017ApJ...850...74K}). \cite{Liu2019XraySO} studied the effect of polar gas on the X-ray spectrum of AGN by considering an X-ray source surrounded by an equatorial disk and a hollow cone perpendicular to the plane of the disk. \cite{Liu2019XraySO} found that the polar gas has a significant effect on the X-ray spectrum of AGN, and argued that the scattered fluorescence line features can act as a potential probe for the kinematics of the polar gas.

    So far, most studies of reprocessed X-ray radiation in AGN have been carried out assuming a torus / disk and no polar component. Thus, adding this polar component in the study of scattered X-ray emission can provide an important look into the kinematics of the polar gas and thus its origin. Since the polar gas is likely optically thin, it would only affect the soft (0.3--5\,keV) part of the X-ray spectrum and not the hard portion ($>5$\,keV). In this paper, we use the ray-tracing simulation software \textsc{RefleX}\footnote{\url{https://www.astro.unige.ch/reflex/}} \citep{RefleX} to simulate the spectra from an AGN with a torus accompanied by a large scale polar filled and hollow cone corresponding to the best fit model of the MIR data. We study the effect of the polar gas on the X-ray spectrum considering multiple inclination angles and column densities.
    
    \begin{figure*}
        \centering
        \includegraphics[scale = 0.14]{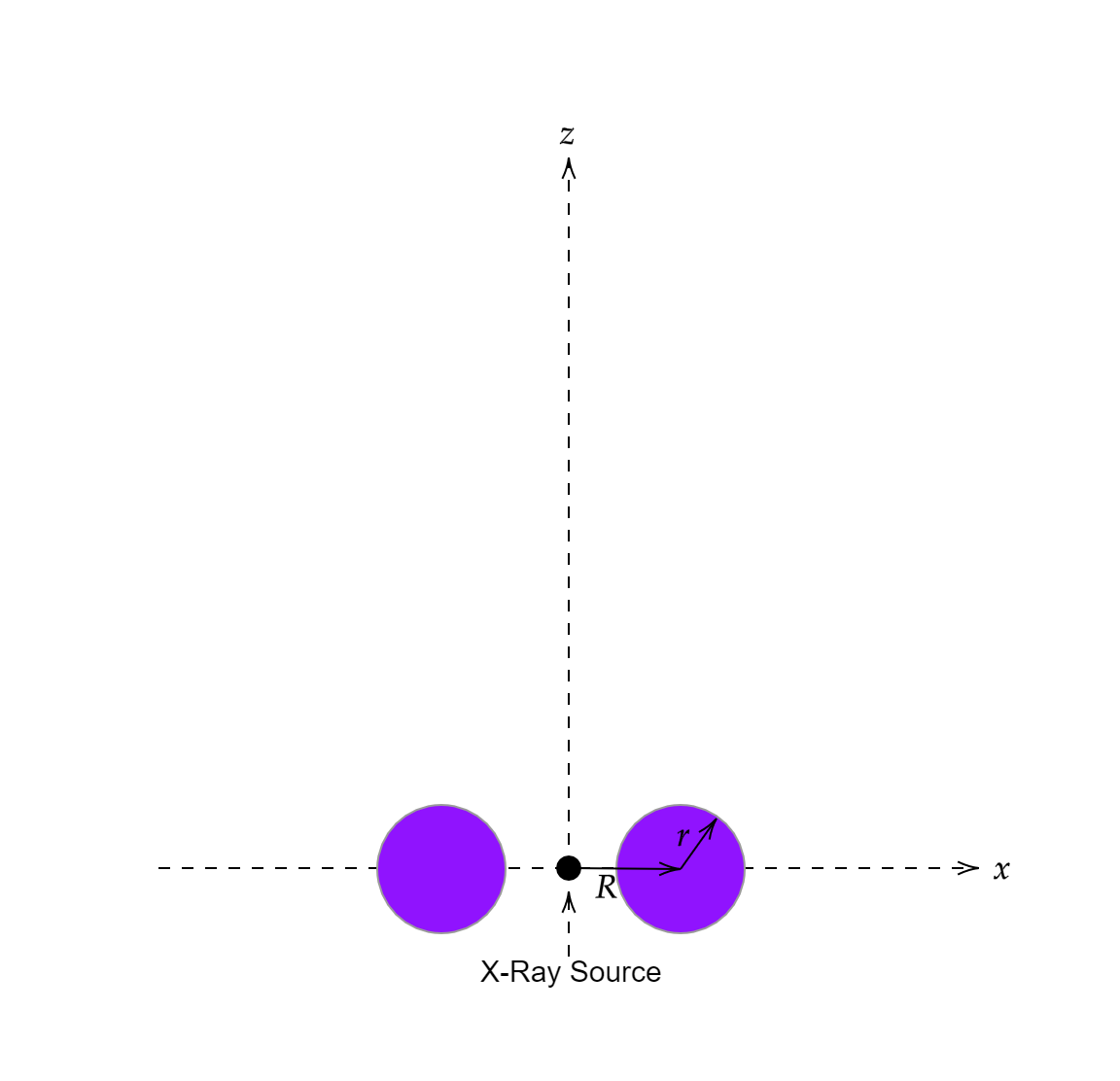}
        \includegraphics[scale = 0.14]{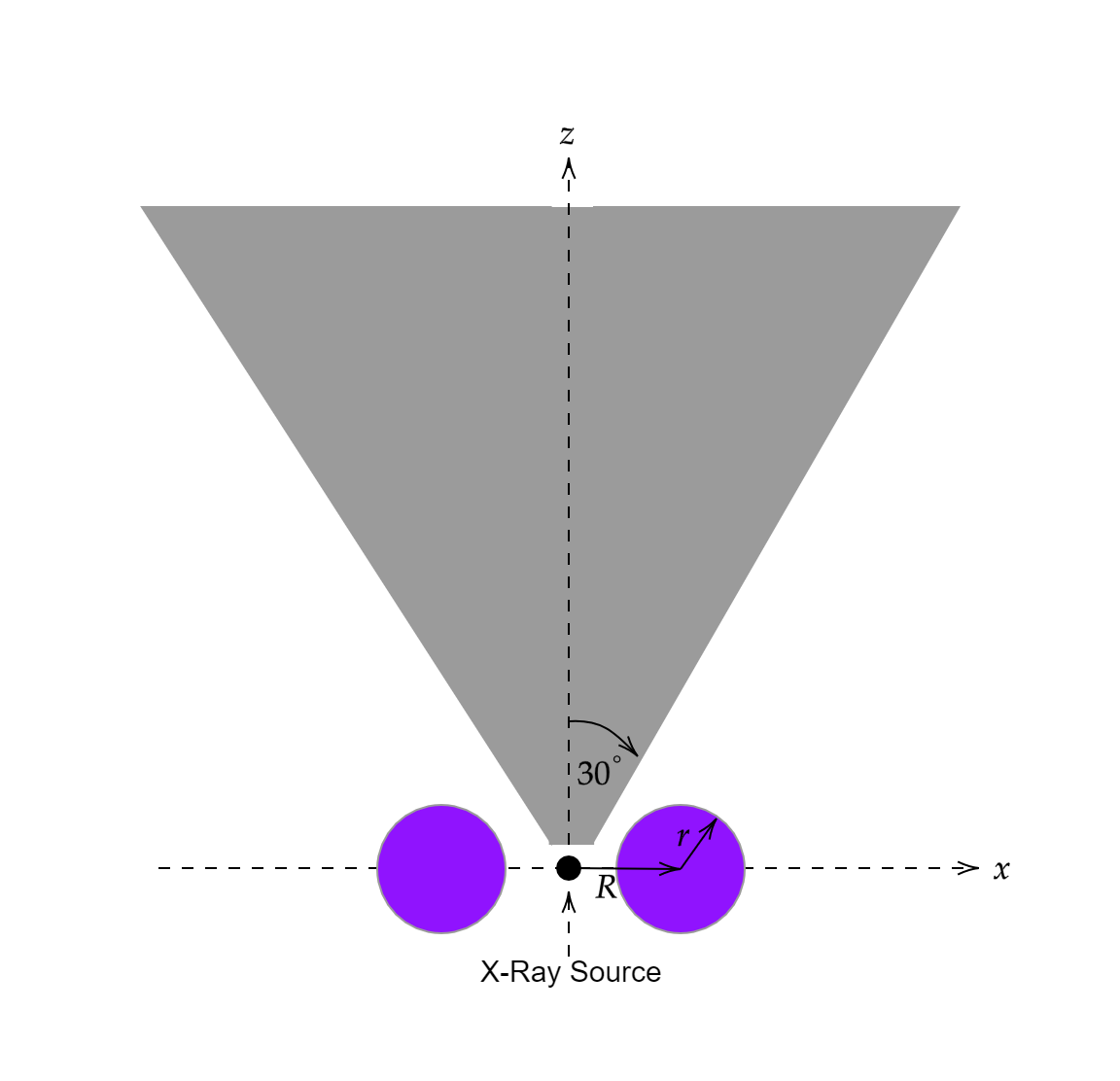}
        \includegraphics[scale = 0.14]{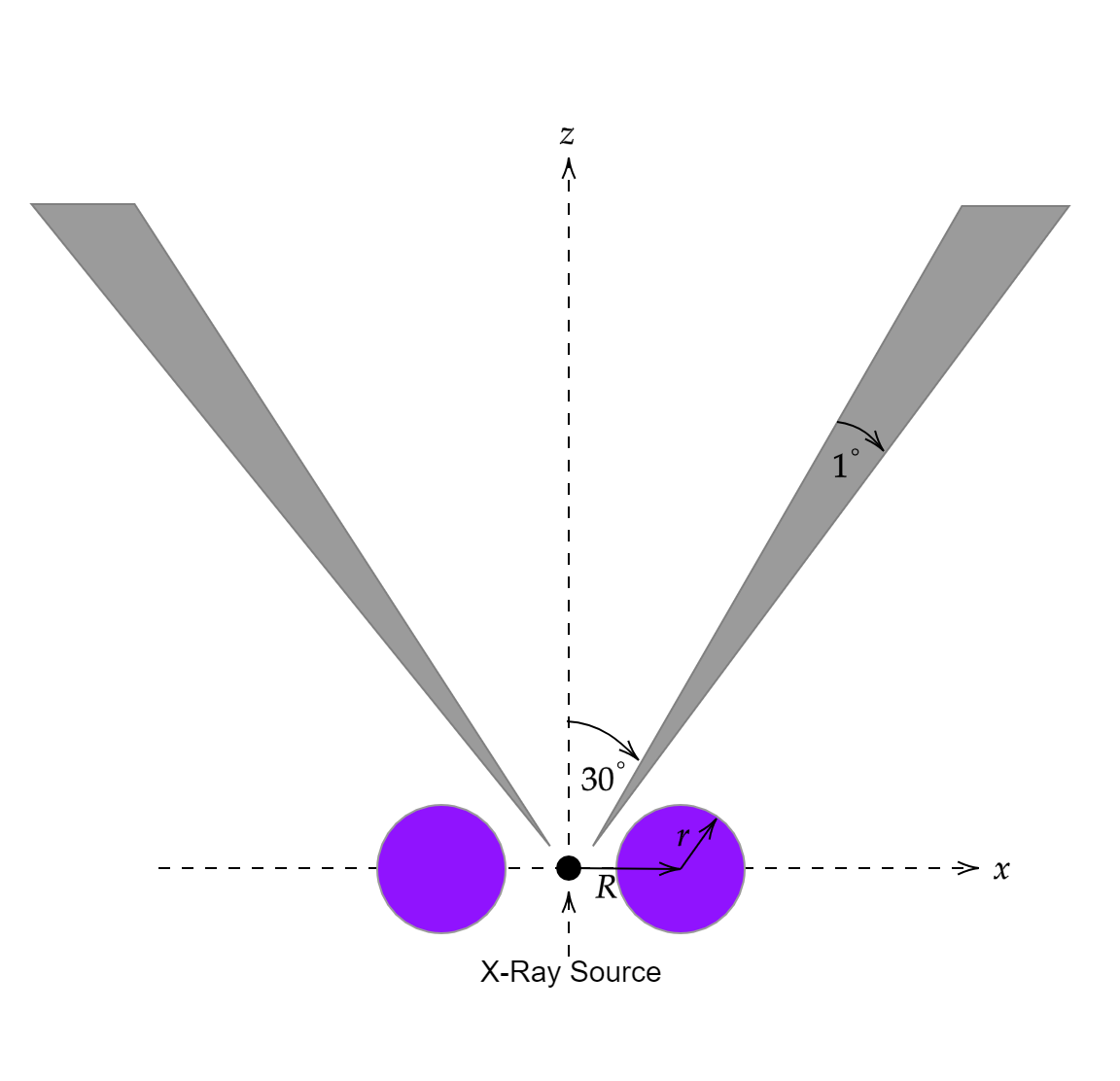}
        \caption{Cross-sections for the three geometries considered. (Left panel) Geometry for torus with radius $r = 1.235$\,pc a distance $R = 1.756$\,pc from the X-ray source. (Middle panel) Geometry for the torus + filled cone. (Right panel) Geometry for the torus + hollow cone. The cones have an opening angle of 30$^{\circ}$ with slant lengths of 40\,pc. The hollow cone has a small angular width of $1^{\circ}$ corresponding to the best fit MIR data for the galaxy Circinus.}
        \label{fig1}
    \end{figure*}

    This paper is laid out as follows. In $\S$2 we give an overview of the parameters and models of our simulations. Then we discuss the results of our simulations in $\S3$. Next, in $\S$4 we lay out the equivalent widths of the spectral lines to be compared to current and future observations from X-ray missions as well as consider observational differences in the resulting spectra of our simulations. We also add many components seen from the Narrow Line Region (NLR) such as a scattered power-law and many photoionised lines seen in obscured AGNs. $\S5$ investigates the difference in the spectra when the opening angle and abundances of the hollow cone is changed. Finally, in $\S$6 we summarize our findings and present our conclusions. 

\section{Simulation Setup}

    \subsection{\textsc{RefleX}: a ray-tracing simulation platform}
    
        In this paper we use the ray-tracing simulation platform \textsc{RefleX} \citep{RefleX} in order to study the emission from this new polar region. \textsc{RefleX} is not the first  ray-tracing code to be used in studying reprocessed radiation from AGN. Indeed some of the first ray-tracing simulation platforms date to the early 1990s with Monte Carlo methods (e.g., \citealp{1991MNRAS.249..352G}). Monte Carlo simulations became instrumental in simulating the X-ray spectra of heavily obscured AGN. They were used, for example, to calculate the equivalent width of the Iron K$\alpha$ complex produced in neutral matter considering various column densities and Iron abundances \citep{2002MNRAS.337..147M}. Newer models, such as \textsc{MYTorus} \citep{10.1111/j.1365-2966.2009.15025.x}, consider a toroidal geometry with arbitrary column densities, while codes such as \cite{2011MNRAS.413.1206B} and \cite{2018ApJ...854...42B} also include varying covering factors. Clumpy torus models have also been developed recently (e.g., \citealp{2014ApJ...787...52L, 2019A&A...629A..16B}). However, most of these models limit the users input in the allowed geometries. Alternatively, \textsc{RefleX} allows the user to specify an X-ray source with different geometries and various spectral shapes in the 0.1\,keV--1\,MeV energy range. After the geometry and spectral shapes are determined, \textsc{RefleX} allows the user to build quasi-arbitrary gas distributions around the X-ray source, with user specified densities. When the simulation is run, \textsc{RefleX} tracks the life of every X-ray photon as it is propagated though the distribution of gas. The photons can undergo several physical processes when propagating through the gas, such as fluorescence, Compton scattering, and Rayleigh scattering. Once a photon escapes the gas and reaches the detector it is collected in energy bins in terms of flux (keV\,cm$^{-2}$\,s$^{-1}$\,keV$^{-1}$), counts (integer number of photons), or photons (cm$^{-2}$\,s$^{-1}$\,keV$^{-1}$), depending on the specifications of the user.
        
        \textsc{RefleX} also allows the user to specify the composition of the material surrounding the X-ray source. This includes setting the gas metallicity, the fraction of Hydrogen in molecular form (H$_{2}$ fraction), as well as the composition of the material which can be set, for example, to follow what was reported by \cite{1989GeCoA..53..197A} or \cite{2003ApJ...591.1220L}. All simulations in $\S$3 are run using the gas composition from \cite{1989GeCoA..53..197A}, solar metallicity, and a H$_{2}$ fraction of 1. In $\S$5 we will consider the composition from \cite{2003ApJ...591.1220L}, as well as an H$_{2}$ fraction of 0.2, showing that no significant difference is expected in the features arising from the polar gas.    
    
    \subsection{Geometries}
    \subsubsection{Torus model}
    
        In order to study the effect polar gas has on the X-ray spectrum of AGN, we implement three simulation setups. The first is a torus with a Compton thick equatorial column density of $N_{\rm H,\, eq} =10^{24.5}$\,cm$^{-2}$ surrounding an isotropic X-ray source emitting a power-law distribution of photons with an exponential cutoff at 200\,keV and a photon index of $\Gamma = 1.9$. This is equivalent to an AGN with no polar component and thus will provide a comparison to our later simulations with the polar component. The distance and size of the torus is characterized by two parameters in \textsc{RefleX}: $R$ and $r$ as seen on the left panel of Figure\,\ref{fig1}. The values of $R$ and $r$ are kept consistent in all the simulations and have values of $R = 1.756$\,pc and $r = 1.235$\,pc, corresponding to the best fit model of the Circinus galaxy (Andoine et al. submitted), leading to a covering fraction of $CF = 0.7$. The gas in the torus and all other geometries is smooth. For the torus (as well as all other geometries), we selected the spectra from four inclination angle ranges: $0^{\circ} \leq i  \leq 5^{\circ}$ ($N_{\text{H}} = 0$\,cm$^{-2}$), $45.6^{\circ} \leq i \leq 50.6^{\circ}$ ($N_{\text{H}} = 10^{22.49}$\,cm$^{-2}$), and $85^{\circ} \leq i \leq 90^{\circ}$ ($N_{\text{H}} \approx 10^{24.50}$\,cm$^{-2}$) where $i$ is the angle measured from the normal to the plane of the torus. The column densities are calculated from the center of the range. The range $45.6^{\circ} \leq i \leq 50.6^{\circ}$ will be referred to as the intermediate case and the ranges $0^{\circ} \leq i \leq 5^{\circ}$ and $85^{\circ} \leq i \leq 90^{\circ}$ will be referred to as the pole-on and edge-on cases respectively. We also ran simulations at an inclination range of $60^{\circ} \leq i \leq 65^{\circ}$, but with no significant change from the $45.6^{\circ} \leq i \leq 50.6^{\circ}$ case. We note in our simulations we use a linear resolution of 5\,eV for $E \leq 10$\,keV and a logarithmic resolution of 0.001\,eV for $E > 10$\,keV   
        
        \begin{figure} 
            \centering
            \includegraphics[scale = 0.55]{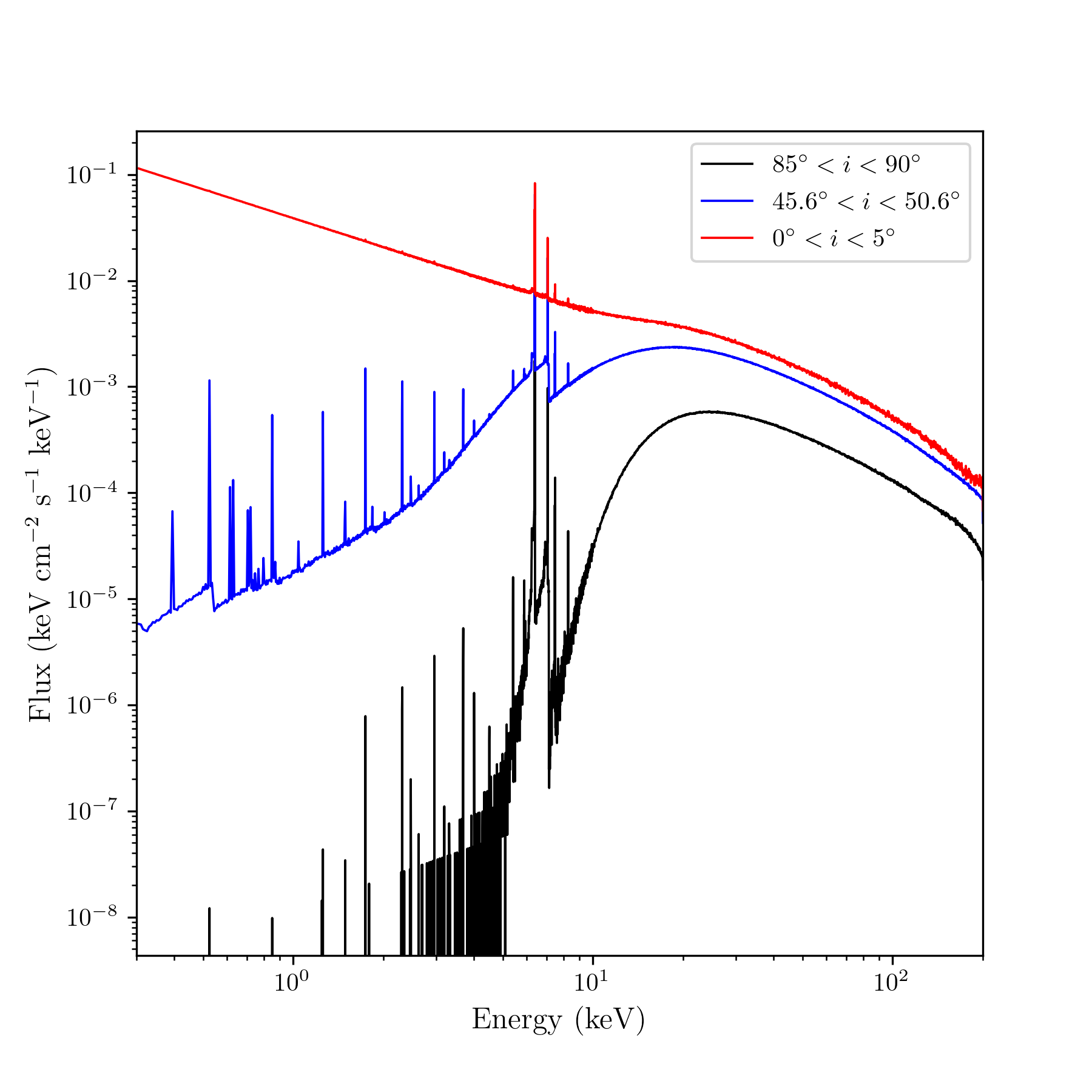}
            \caption{Simulated 0.3$-$200\,keV spectrum obtained considering a torus with an equatorial column density $N_{\text{H, eq}} = 10^{24.5}$\,cm$^{-2}$ at different inclination angles. All the fluxes are normalized as discussed in section 3.1. The edge-on case (black line) has a high photoelectric cutoff due to the strong absorption from the torus. The intermediate case (blue line) displays many fluorescence lines in the soft portion of the spectrum due to the decreased column density. The pole-on inclination angle just shows strong continuum due to the complete lack of line of sight absorbing material along with the Fe K$\alpha$ complex.}
            \label{fig2}
        \end{figure}
        
        \begin{figure*}
            \ContinuedFloat*
            \includegraphics[scale = 0.384]{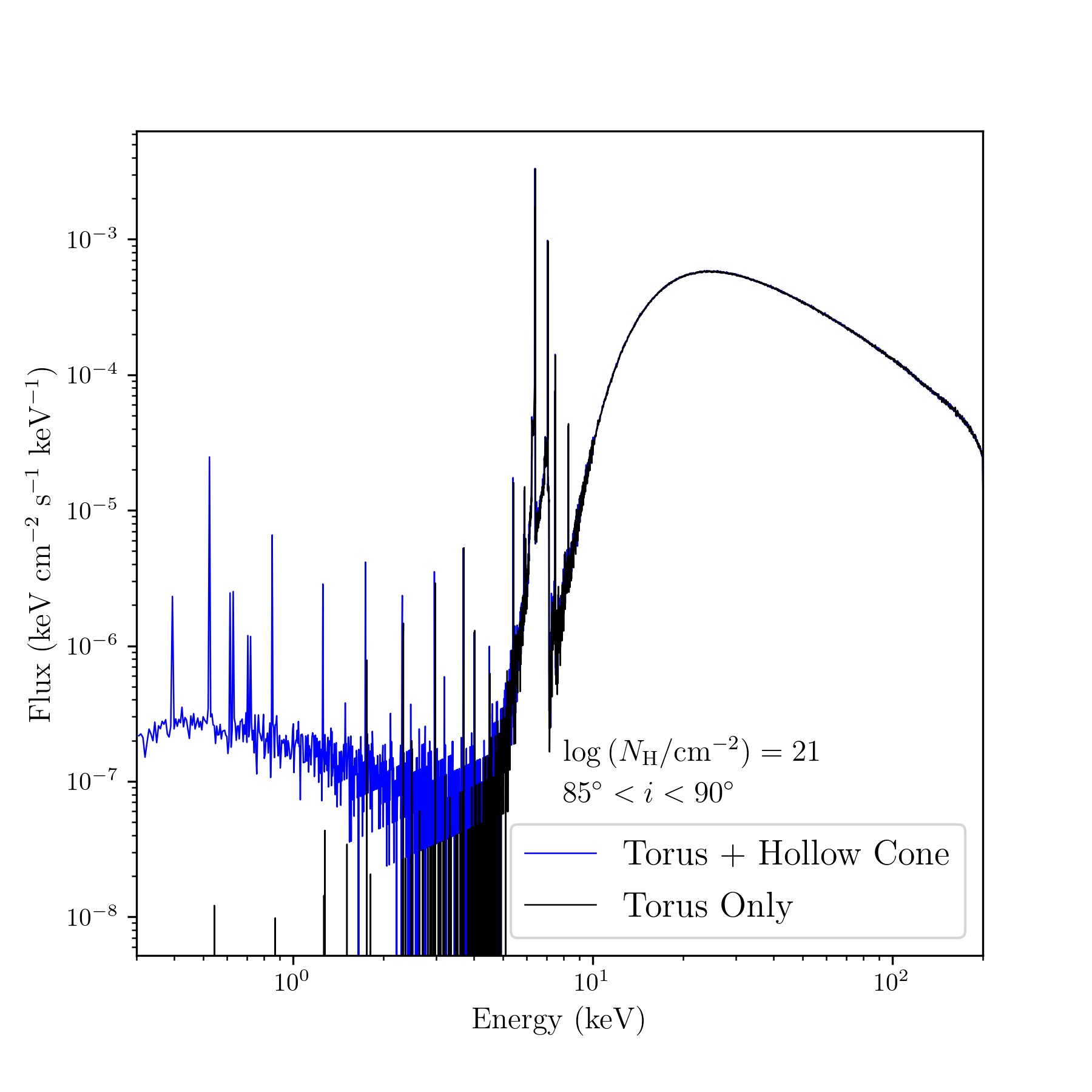}
            \includegraphics[scale = 0.384]{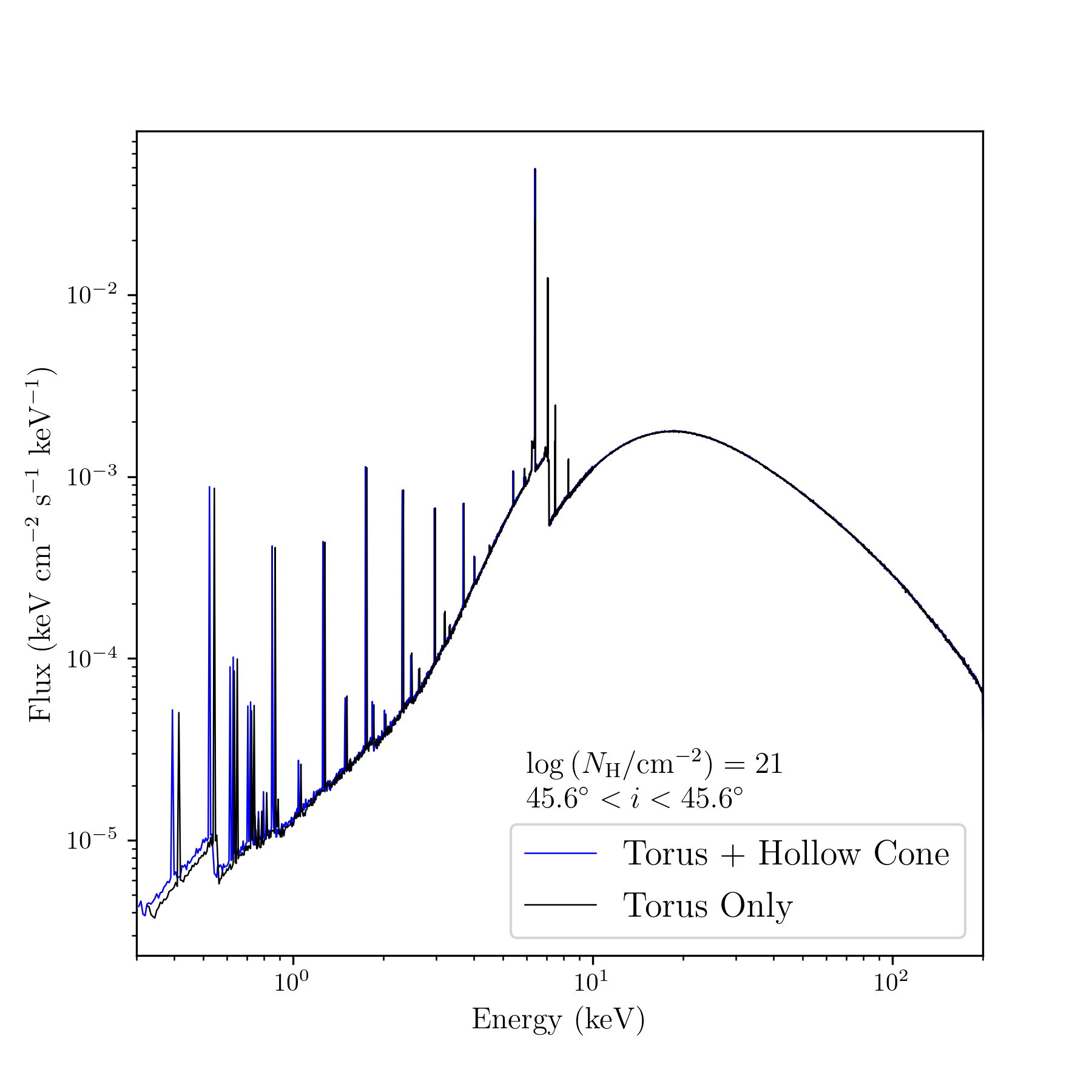}
            \includegraphics[scale = 0.384]{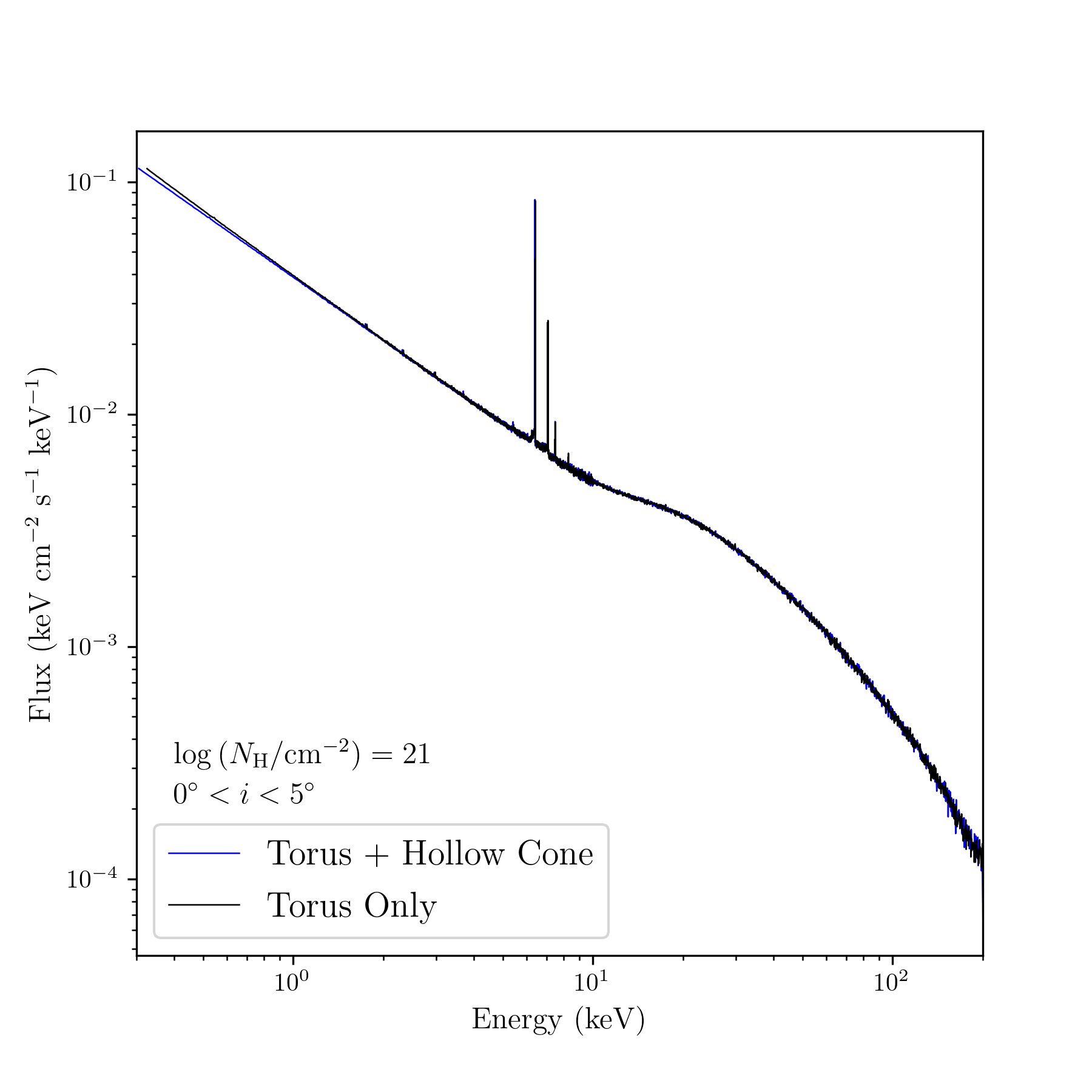}
            \caption{\label{fig3a}Simulated spectra for a torus + hollow cone geometry (shown in blue) along with the torus only simulations (shown in black). Moving from the left panel to the right panel, inclination angle decreases as the slant column density of the hollow cone remains constant ($\log{N_{\text{H}}}\big/\text{cm}^{-2} = 21$). The torus only simulations shown are linearly shifted to higher energies by 0.02\,keV to better view the differences between spectral lines in the soft spectrum. The largest difference between the torus + hollow cone simulations and the previous torus simulations comes at the edge-on case where the previous torus simulations contain a photoelectric cutoff (left panel). The intermediate case (middle panel) shows little deviation from the torus only case. This is also the case for the pole-on case (right panel) due to the continuum domination.}
        \end{figure*}
        \begin{figure*}
            \ContinuedFloat
            \includegraphics[scale = 0.384]{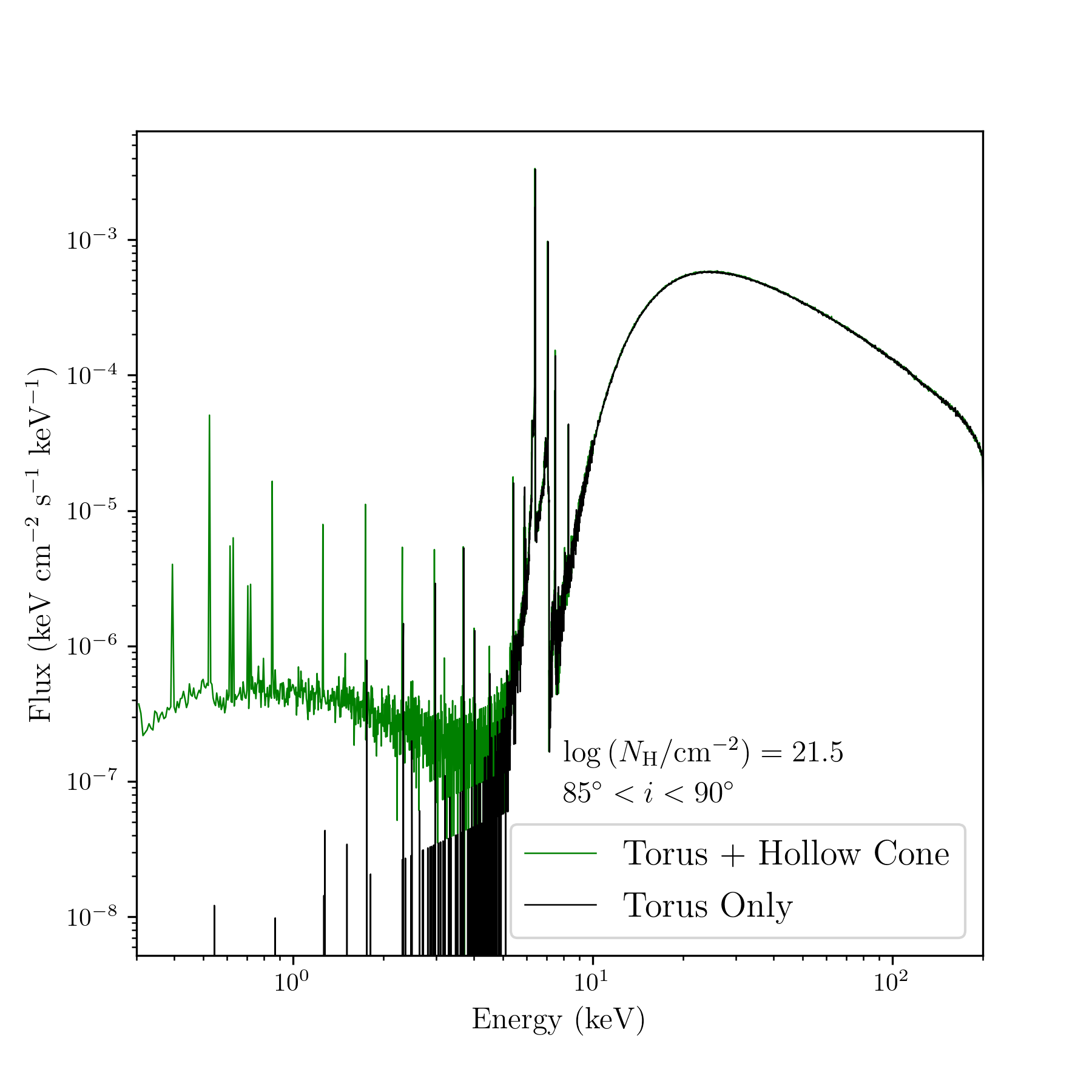}
            \includegraphics[scale = 0.384]{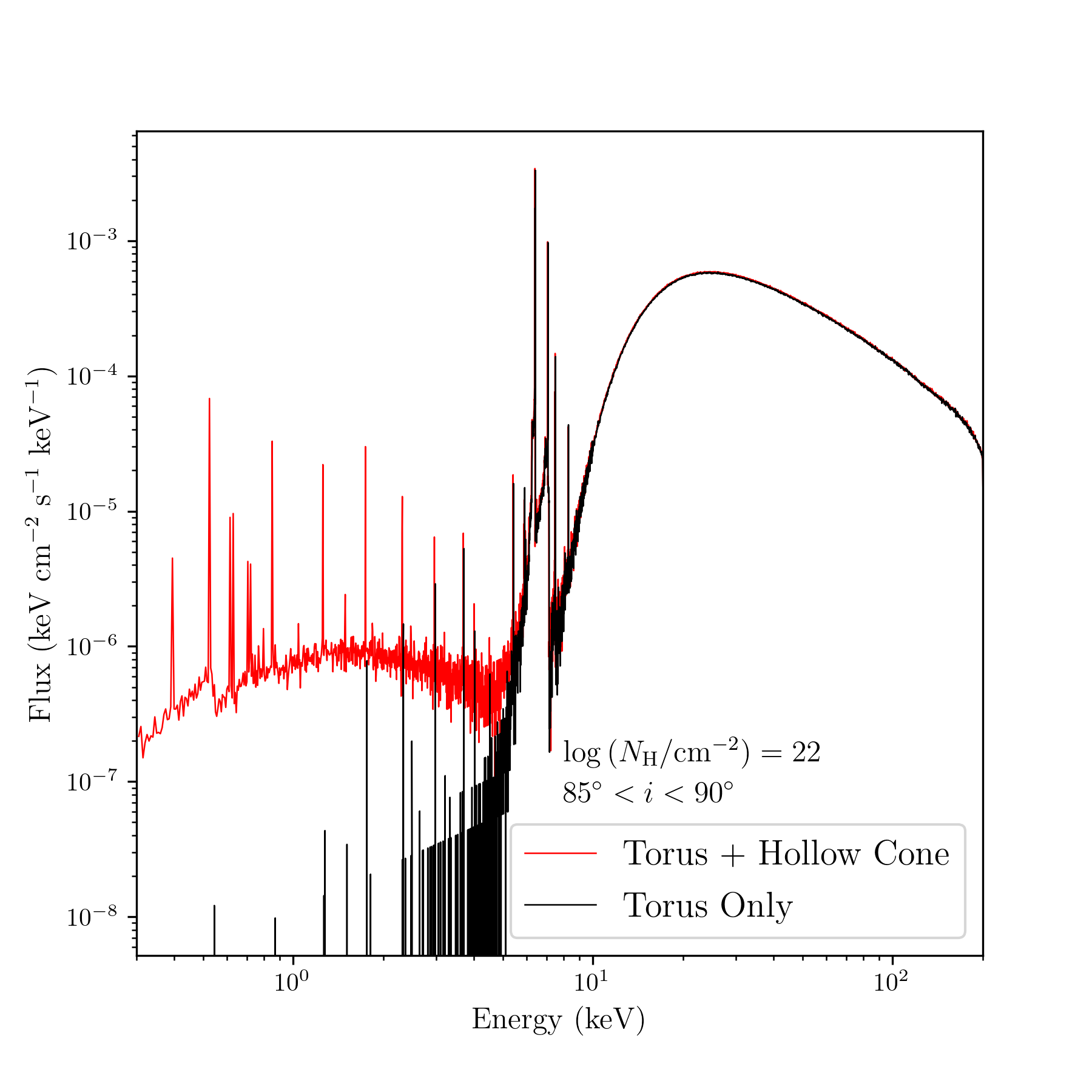}
            \includegraphics[scale = 0.384]{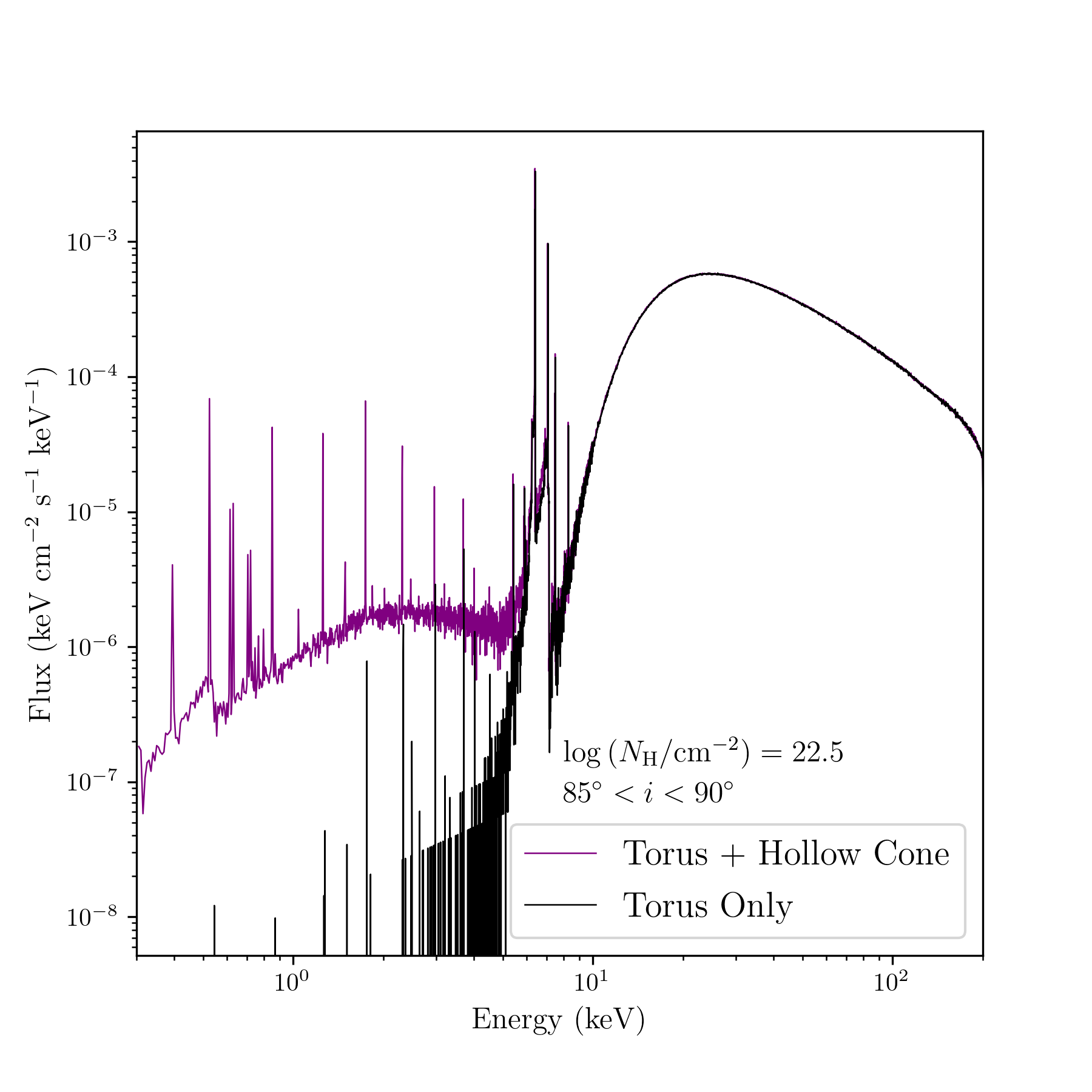}
            \caption{\label{fig3b}Same as top, however, moving from the left panel to the right panel represents simulations with constant inclination angle ($85^{\circ} \leq i \leq 90^{\circ}$) and increasing slant column densities for the polar component. The black line again represents the previous torus only simulation for the edge-on case. As the density of the polar component increases, the flux when compared with the torus only simulations increase as there is now more gas for photons to interact with.} 
        \end{figure*}
    
    \subsubsection{Torus + Hollow/Filled Cone Model}
    
         The geometry for the torus along with the hollow cone can be seen on the right panel of Figure\,\ref{fig1}. The slant length of the cone was chosen to be 40\,pc with an opening angle of $\alpha = 30^{\circ}$ and small angular width of 1$^{\circ}$, consistent with the best MIR model for the Circinus galaxy \citep{10.1093/mnras/stx2227, 10.1093/mnras/stz220}. The same three inclination angles are considered as in the torus-only model. However, to investigate the effect of the density of the polar gas on the X-ray spectrum, four slant column densities\footnote{the column density as seen looking along the slant length of the cone} are considered for the hollow cone at each inclination angle. These are $\log{(N_{\rm H}\big /\text{\,cm}^{-2})} = 21, 21.5, 22,$ and $22.5$. The cone starts at a height of 0.1\,pc above the X-ray source, corresponding to the sublimation radius of dust in AGN (e.g., \citealp{2007A&A...476..713K}). The setup for the simulations including the torus along with the filled cone can be seen in the middle panel of Figure\,\ref{fig1}. The filled cone is equivalent to the hollow cone except for the fact that the outer wall of the hollow cone is removed and the inner portion of the cone is filled with gas. The same three inclination angles and column densities are considered for both the hollow and filled cone. It should be noted that it is unlikely that the polar gas has such a filled geometry in AGN. Otherwise, we would expect a large fraction of type 1 AGN spectra to show obscured X-ray spectral features, which is not the case \citep{Ricci2017, 2017ApJ...850...74K}. The material in these cones is also thought to be dusty (i.e., \citealp{10.1093/mnras/stx2227, 10.1093/mnras/stz220}) and thus must have a low level of ionisation. Therefore, we assume neutral material for the cones.  

    \begin{figure*}
        \centering
        \includegraphics[scale = 0.35]{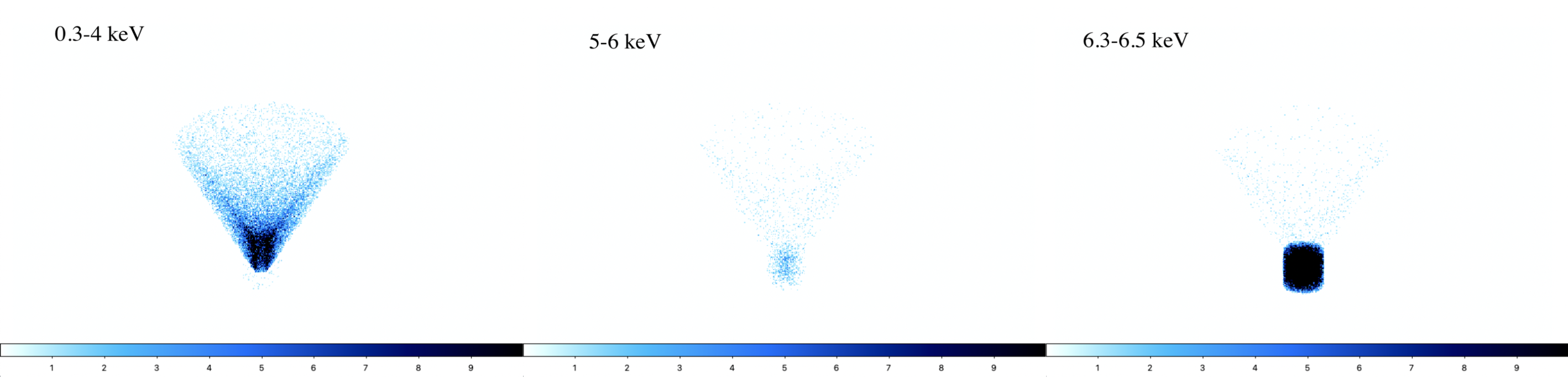}
        \caption{Images of the torus + hollow cone simulations for the edge-on case at the highest polar gas column density ($85^{\circ} \leq i \leq 90^{\circ}$, $\log{N_{\text{H}}}\big/\text{cm}^{-2} = 22.5$). The colour bars represent the number of total interactions (scattering + fluorescence) undergone in that particular pixel. (Left panel) Image showing the most interactions in the 0.3--4\,keV range takes place is in the polar component. (Middle panel) 5--6\,keV range which is mostly devoid of interactions due to the fact that this range is dominated by continuum. (Right panel) 6.3--6.5\,keV range which is dominated by the torus due to the production of the Iron K$\alpha$ line at $\sim6.4$\,keV.}
        \label{fig4}
    \end{figure*}
    
    \begin{figure}
        \centering
        \includegraphics[scale = 0.55]{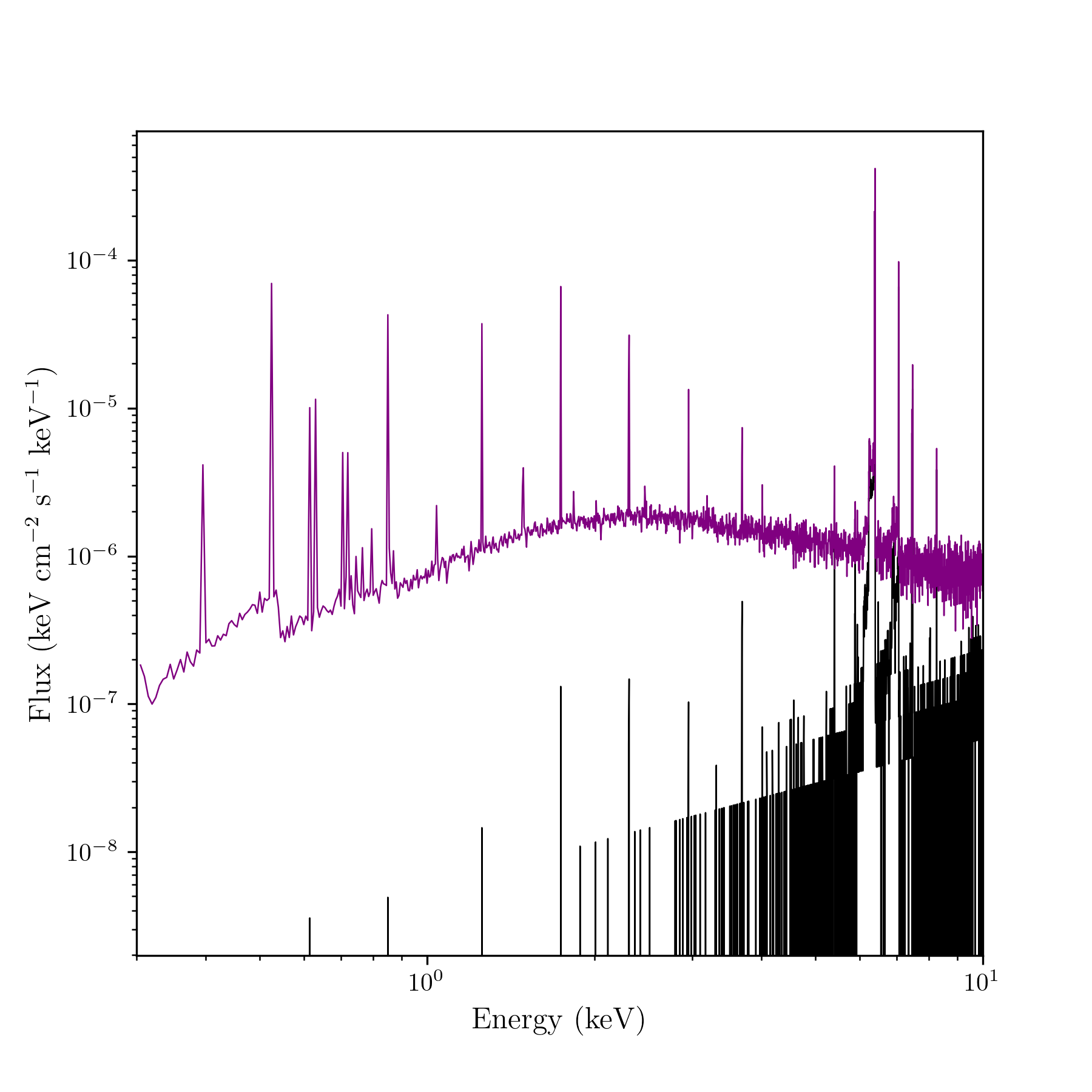}
        \caption{Simulation showing the result of increasing the equatorial column density of the torus to $N_{\text{H, eq}} = 10^{25}$\,cm$^{-2}$ for the edge-on inclination range and the highest polar gas column density. The torus only component (black) shows very poor photon transmission due to the increased column density allowing most of the emission to stem from the polar component in the total spectrum (purple).}
        \label{fig4.5}
    \end{figure}
    
\section{Results}

    \subsection{Torus}
    
        First we present the results for the torus only geometry as to provide a comparison for our later simulations with the polar component. The results for the torus simulations at each inclination angle are shown in Figure\,\ref{fig2}. The the edge-on case ($85^{\circ} \leq i \leq 90^{\circ}$) is characterized by a high photoelectric absorption due to the large column density in the torus and a more prominent Iron K$\alpha$ line at $\sim$6.4\,keV which is also seen at all other inclination ranges. The Compton shoulder is also visible at $\sim$ 6.2--6.4\,keV. The simulation run at the intermediate inclination angle ($45.6^{\circ} \leq i \leq 50.6^{\circ}$) has a much lower column density ($N_{\rm{H}} = 10^{22.49}$\,cm$^{-2}$) which allows for the detection of more fluorescence lines in the soft portion of the spectrum, due to the decreased importance of photoelectric absorption. The fluorescence lines implemented in \textsc{RefleX} can be found in Appendix A of \cite{RefleX}. The hard portion of the spectrum in all simulations shows a reflection hump peaking $\sim$ 20--30 keV. The pole-on case shows no fluorescence lines in the soft band as the line of sight is no longer obscured by the torus and the spectral lines are washed out by the dominating X-ray continuum.
    
    \subsection{Torus + Hollow Cone}
    
        The results for all the torus + hollow cone simulations for the different inclination angles and polar gas column densities are shown in Figures\,\ref{fig3a} and \ref{fig3b}. Moving from left to right, Figure\,\ref{fig3a} shows how the X-ray spectra are affected by the polar gas as different inclination angles are considered while keeping column density constant (in this case, the column density shown is $\log{N_{\mathrm{H}}}\big/\text{cm}^{-2} = 21$). Figure\,\ref{fig3b} shows how the X-ray spectra are affected by the polar gas as different slant column densities are considered while keeping inclination angle constant. The inclination angle shown in Figure\,\ref{fig3b} is $85^{\circ} \leq i \leq 90^{\circ}$. The black spectra shown in each panel shows the torus only simulation at the same inclination angle. This gives the ability to clearly distinguish which parts of the X-ray spectrum are most affected by adding a polar component to the simulations. The simulations with a hollow cone clearly show the most deviations from the torus at the edge-on case in the 0.3--4\,keV portion of the spectrum at the edge-on inclination range. The appearance of the polar gas allows many strong fluorescence line features to dominate the 0.3--4\,keV range due to its optically thin nature. The strongest of these lines being O, Ne, Mg, and Si, with the O line dominating in equivalent width when the polar gas is at its highest column density and the largest inclination angle ($\log{N_{\mathrm{H}}}\big/\text{cm}^{-2} = 22.5$, $85^{\circ} \leq i \leq 90^{\circ}$; see Table \ref{table1}). The flux of these fluorescence lines increase with the density of the polar gas. However, for the intermediate and pole-on cases, little deviation from the torus model occurs. The deviations that do occur in the intermediate case only appears in the very soft (0.3--1.2\,keV) portion of the spectrum.
        
        In addition to these simulations, we also considered how the new polar component would be effected if the torus considered had a much higher equatorial column density, such as $N_{\text{H, eq}} = 10^{25}$\,cm$^{-2}$. The results of this for the edge-on case can be seen in Figure\,\ref{fig4.5} with the torus component shown in black and the torus + hollow cone shown in purple. In this case the X-ray emission is much weaker, due to the higher column density. This allows us to attribute most of the scattered radiation below $\sim$ 6\,keV to the polar component.

        \subsubsection{Imaging the system}
        
            To better understand the morphology of the torus + hollow cone simulations, we took advantage of a feature of \textsc{RefleX} which allows the user to create simulated images. The user inputs the location of a detector, its aperture size, field of view, number of desired pixels, and creates an image. The user can also select the energy range of the photons which are collected. Using this feature, we created three images of the simulations of a torus + hollow cone at the largest inclination angle and highest polar gas density [$85^{\circ} \leq i \leq 90^{\circ}$, $\log{(N_{\text{H}}\big /\text{\,cm$^{-2}$}}) = 22.5$]. We selected three energy ranges to create the images: 0.3--4\,keV, 5--6\,keV, and 6.3--6.5\,keV. In addition to this, we isolated only photons which undergo at least one interaction (scattering or fluorescence) in order to remove the direct emission from the central source. These three images are shown in Figure\,\ref{fig4}. The left panel of Figure\,\ref{fig4} shows a large amount of interactions due to the fluorescent line features in the soft X-rays occurring in the polar region as expected due to the low density nature of the polar gas. The middle panel shows the range from 5--6\,keV which contains less interactions than the other ranges due to the fact that this range is dominated by the scattered continuum and no strong fluorescent line is present. In this range there is emission from both regions, however, more interactions occur in the torus due to its higher density. Finally, the right panel shows the 6.3--6.5\,keV range. This range contains fluorescent emission such as the Iron K$\alpha$ line, and a dominating torus component due to the fact that the hollow cone is not dense enough to scatter those high energy photons. 
            
        \begin{figure}
            \centering
            \includegraphics[scale = 0.55]{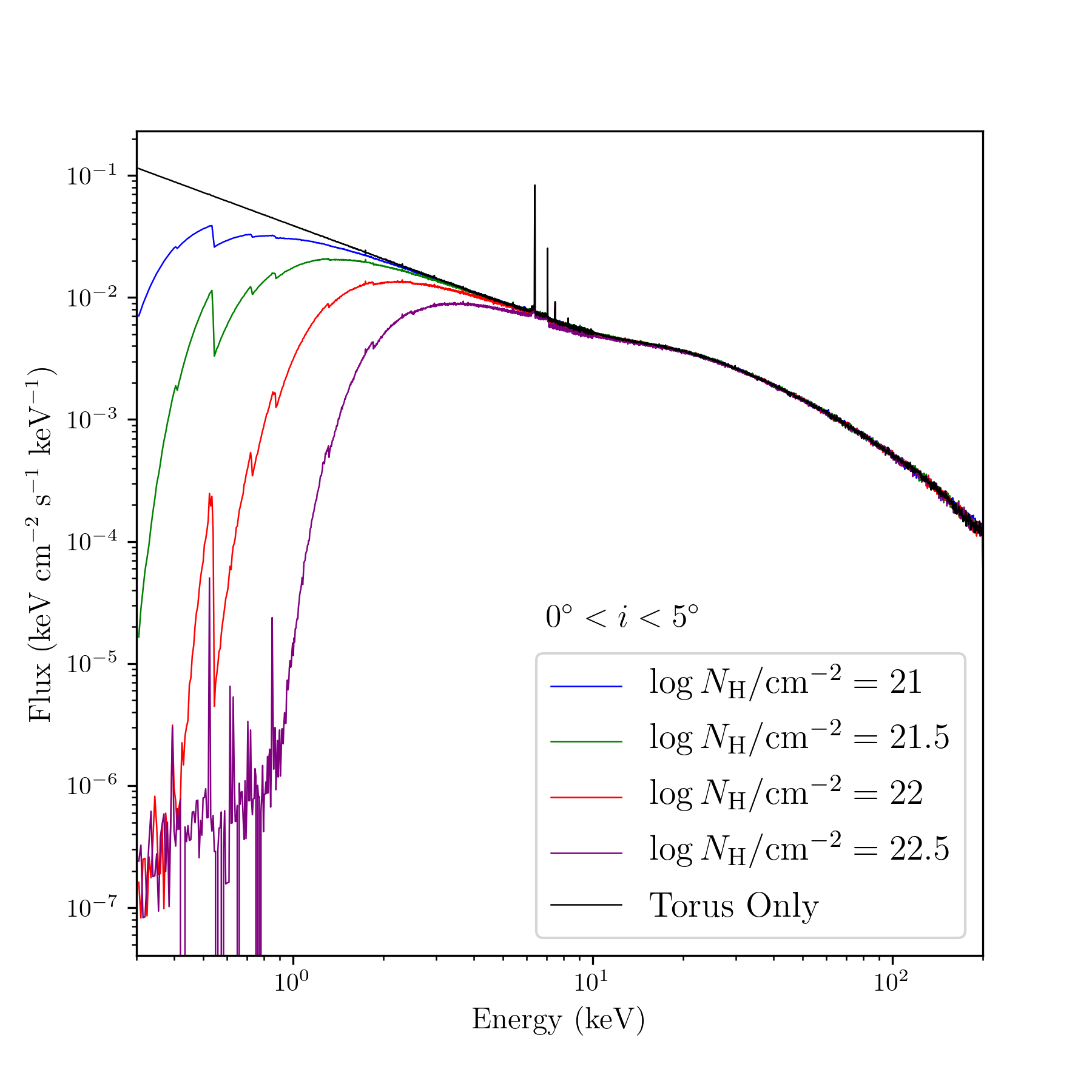}
            \caption{This figure shows the pole-on spectra for the torus + filled cone for the four slant column densities considered. All the column densities show absorption in the soft portion of the spectrum which shows that this geometry for the polar component is unlikely, as Type 1 AGN spectra would be obscured if this was the true geometry of the polar component.}
            \label{figCone}
        \end{figure}
        
    \subsection{Torus + Filled Cone}
    
        When considering the geometry of a torus + filled cone, the spectra show many similarities to the simulations with the torus + hollow cone. These similarities are a large increase in flux and fluorescence line features in the soft portion of the spectrum while the hard portion of the spectrum remains the same as the simulations for a torus-only geometry, with a large reflection hump, Fe K$\alpha$ line, and continuum domination. In addition to this, the intermediate case shows little deviation in flux from the torus-only case. However, the filled cone deviates from the hollow cone in a few important ways. First, the flux increase in the soft portion of the spectrum is not as large as in the hollow cone case. This is due to the fact that in this geometry it is more likely for the photons to be absorbed as there is more gas to interact with. Also, Figure\,\ref{figCone} shows at the pole-on case, the filled cone does not show only continuum, as the X-ray source is obscured by the polar gas, unlike the pole-on case for the hollow cone. This allows for some line features to appear in the soft spectrum for the pole-on case. It was noted earlier that this geometry is unlikely for the polar component as many Type 1 AGN spectra would be obscured. Simulations were also performed considering the same total number of atoms in the filled cone as in the hollow cone and showed that the spectra still showed a clear photoelectric cutoff at low energies, giving further evidence that this geometry is unlikely for the polar component.  
        
        \begin{figure}
            \centering
            \includegraphics[scale = 0.55]{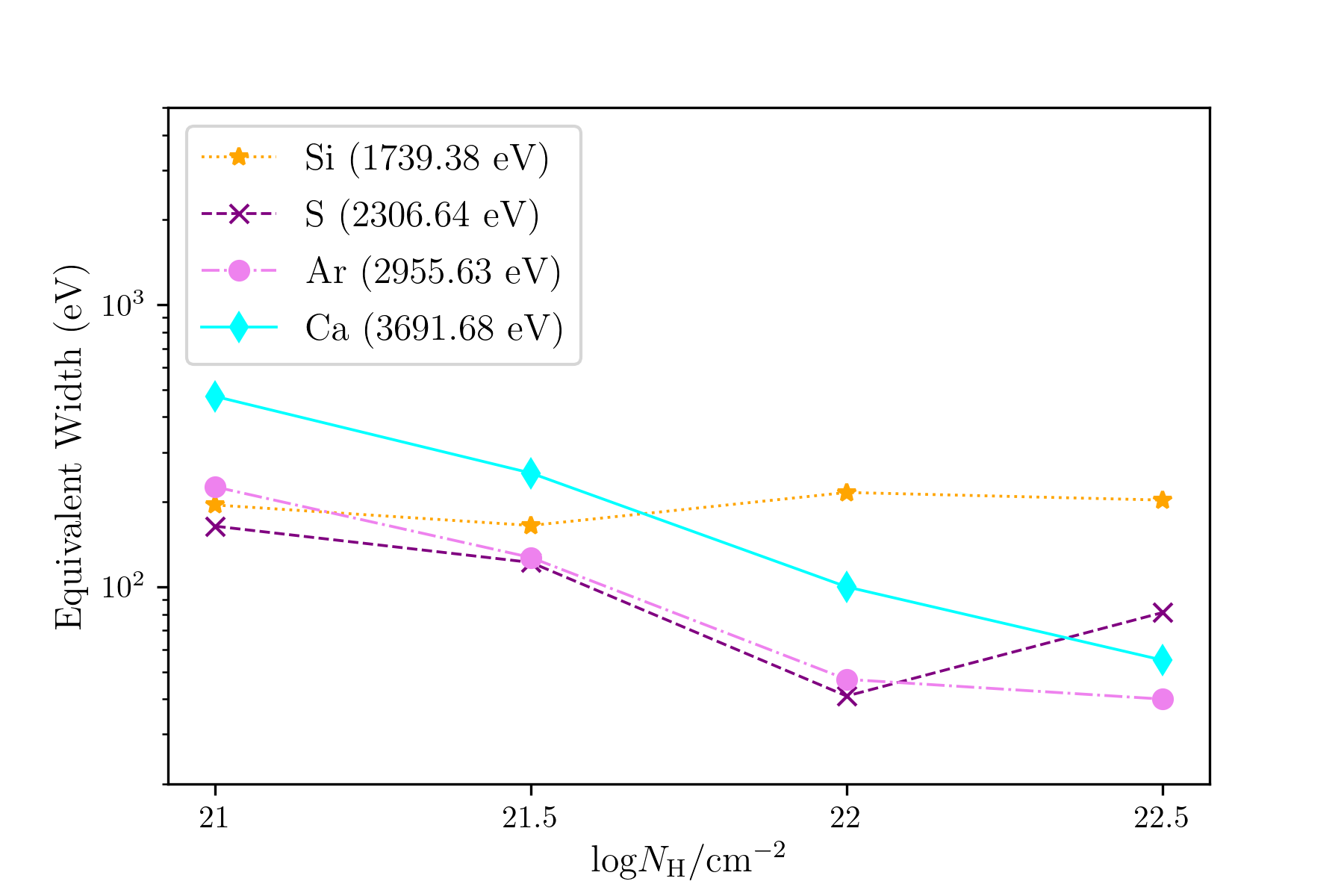}
            \includegraphics[scale = 0.55]{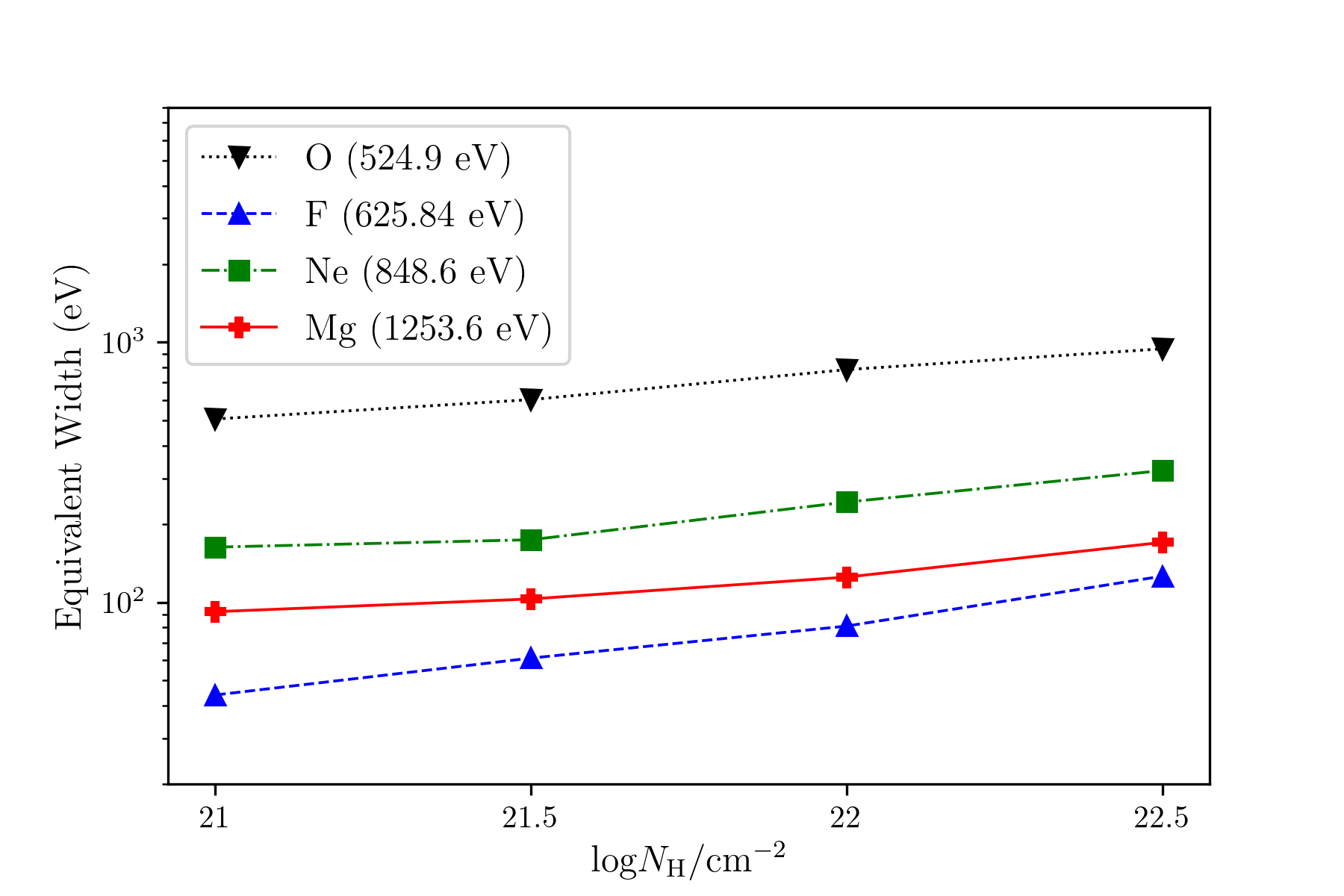}
            \caption{Equivalent widths of spectral lines for a torus + hollow cone as a function of polar gas column density for the edge-on simulations ($85^{\circ} \leq i \leq 90^{\circ}$). (Top panel) EWs for the Si, S, Ar, and Ca lines as a function of polar gas column density. (Bottom panel) Same as top but for the O, F, Ne, and Mg lines. These plots show a clear energy dependence with the EWs of lines of lower energy (O, F, Ne, Mg, Si) increasing while lines with higher energy (S, Ar, Mg) decrease in EW due to the increasing continuum at higher energies. The equivalent widths for all simulations are reported in Tables \ref{table1}-\ref{table6} in Appendix A.}
            \label{fig6}
        \end{figure} 
        
    \section{Detecting Polar Gas in Obscured AGN}
        
        \subsection{Equivalent Width of Spectral Lines}
    
            We discuss here the equivalent widths of the strongest spectral lines in the 0.3$-$4\,keV range for the torus + hollow cone simulations (see Figures\,\ref{fig3a} and \ref{fig3b}). The simulations for the pole-on case ($0^{\circ} \leq i \leq 5^{\circ}$) were ignored since the continuum is not suppressed and thus contrast between the lines and the continuum is not enough for detection. Therefore, these equivalent width measurements are achievable only in an absorbed AGN. The equivalent widths for these simulations are reported in Tables \ref{table1}-\ref{table6} in Appendix A. We also note that, when considering a galactic absorption of 10$^{21}$\,cm$^{-2}$, no significant difference in the lines occurred. However, significant deviations in ISM abundances may cause challenges in attributing these EWs to any particular geometry.     
        
            A plot of the equivalent widths (EWs) as a function of polar gas density of the chosen spectral lines for a torus + hollow cone for the edge-on case is shown in Figure\,\ref{fig6}. The O, F, Ne, Mg, and Si lines all show an increase in equivalent widths as the column density of the polar gas increases, with the O line increasing by the largest amount ($\Delta\text{EW}=$ 437\,eV) followed by the Ne ($\Delta\text{EW}=$ 158\,eV) and Florine ($\Delta\text{EW}=$ 82\,eV) lines. This increase is due to the fact that, as the density of the polar gas increases, there is more material for which to scatter the soft X-rays and create larger fluorescence lines. An interesting feature seen in this plot is that the equivalent widths of the S, Ar and Ca lines decrease with the polar gas column density. To investigate this further, we noted that the flux of the continuum around these lines increases at a larger rate than the line fluxes themselves as we increase the column density. This explains the behaviour of those two lines, since the equivalent width is proportional to the ratio between the flux of the line and that of the continuum.   

        \subsection{Observational Differences in Simulations}
        
            \begin{figure}
                \centering
                \includegraphics[scale = 0.55]{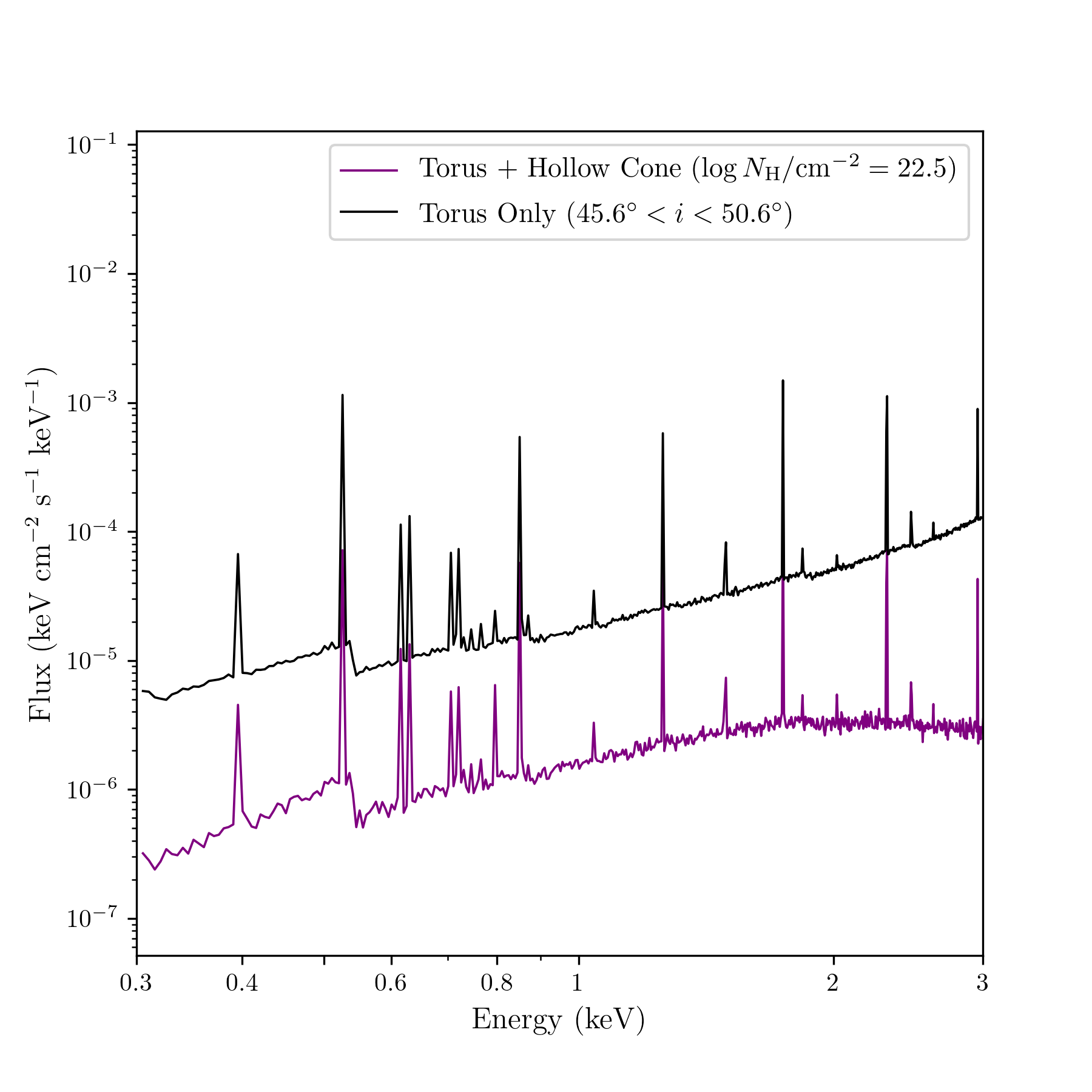}
                \caption{Flux in the soft X-ray band for the edge-on case of a torus + hollow cone (purple) with the torus-only (black) intermediate case ($45.6^{\circ} \leq i \leq 50.6^{\circ}$). This shows the comparison of the slopes of the power-laws in the soft X-ray bands for these two geometries and inclinations with the intermediate torus only case having a power-law slope of $\Gamma = -1.63$ and the edge-on torus + hollow cone having a slope of $\Gamma = -0.8$.}
                \label{fig8.5}
            \end{figure}
        
            \begin{figure}
                \centering
                \includegraphics[scale = 0.365]{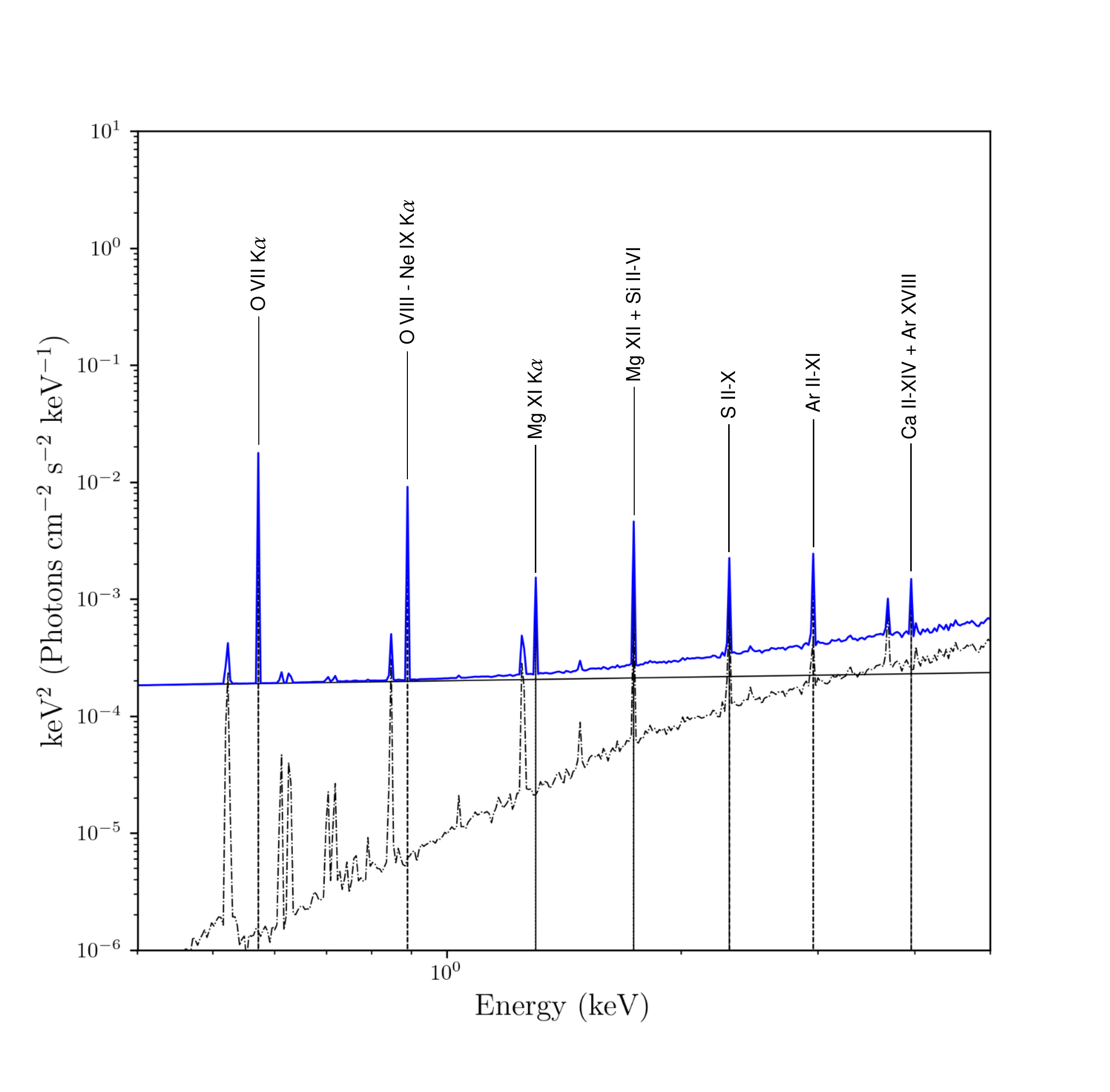}
                \caption{Plot showing the additional components which will most likely be observed stemming from the NLR added to the edge-on spectrum for a torus + hollow cone at the highest polar gas column density considered (original simulation shown as the dash-dotted black line). This includes a power-law model with 0.1\% of the flux of the primary continuum (solid black line), as well as many photoionized lines seen in the NLR such as the ones from the soft X-ray spectrum of NGC 3393 (see Table 4 in \citealp{2006A&A...448..499B}), as well as the Circinus galaxy (see Tables 1, A1 of \citealp{2001ApJ...546L..13S, 2014ApJ...791...81A}, Andonie et al. submitted, respectively). These photoionized lines are plotted as dashed lines in the figure and are labeled with their transitions.}
                \label{fig10.5}
            \end{figure}
            
            \begin{figure}
                \centering
                \includegraphics[scale = 0.42]{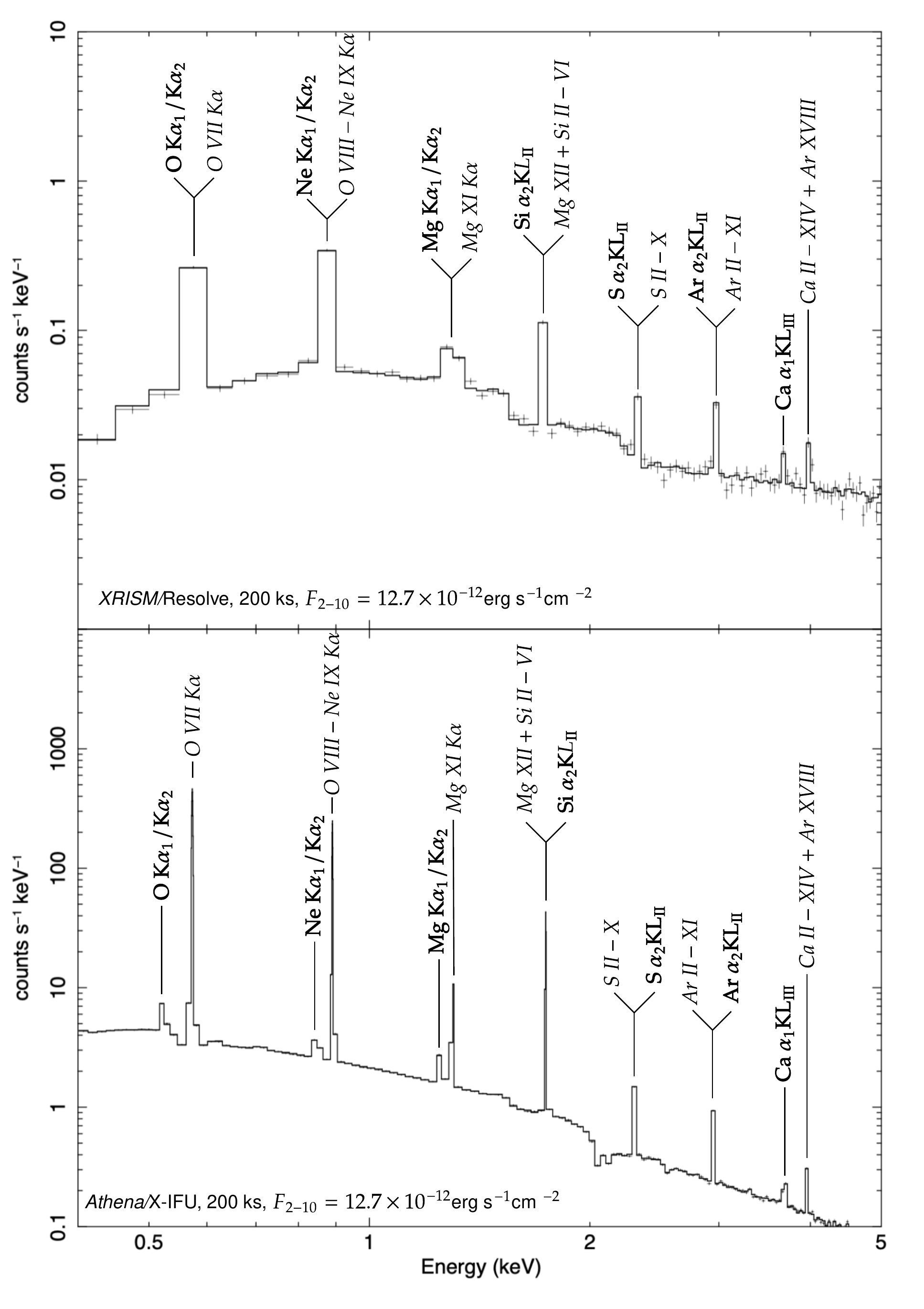}
                \caption{(Top panel) Spectrum with a torus + hollow cone for the edge-on case ($85^{\circ} \leq i \leq 90^{\circ}$) with the highest polar gas column density ($\log{N_{\mathrm{H}}}\big/\text{cm}^{-2}$ = 22.5) with \textit{XRISM} \citep{xrismscienceteam2020science} response and background files. (Bottom panel) Same as top but with \textit{Athena} \citep{2016SPIE.9905E..2FB} response and background files (Barret et al. 2021 in prep.). For both simulations, we used an exposure time of 200\,ks and considered a $2-10$\,keV flux of $F_{2-10} = 12.7\times10^{-12}$\,erg\,s$^{-1}$\,cm$^{-2}$, as in the Circinus galaxy.}
                \label{figres}
            \end{figure}
            
            \begin{figure*}
                \centering
                \includegraphics[scale = 0.55]{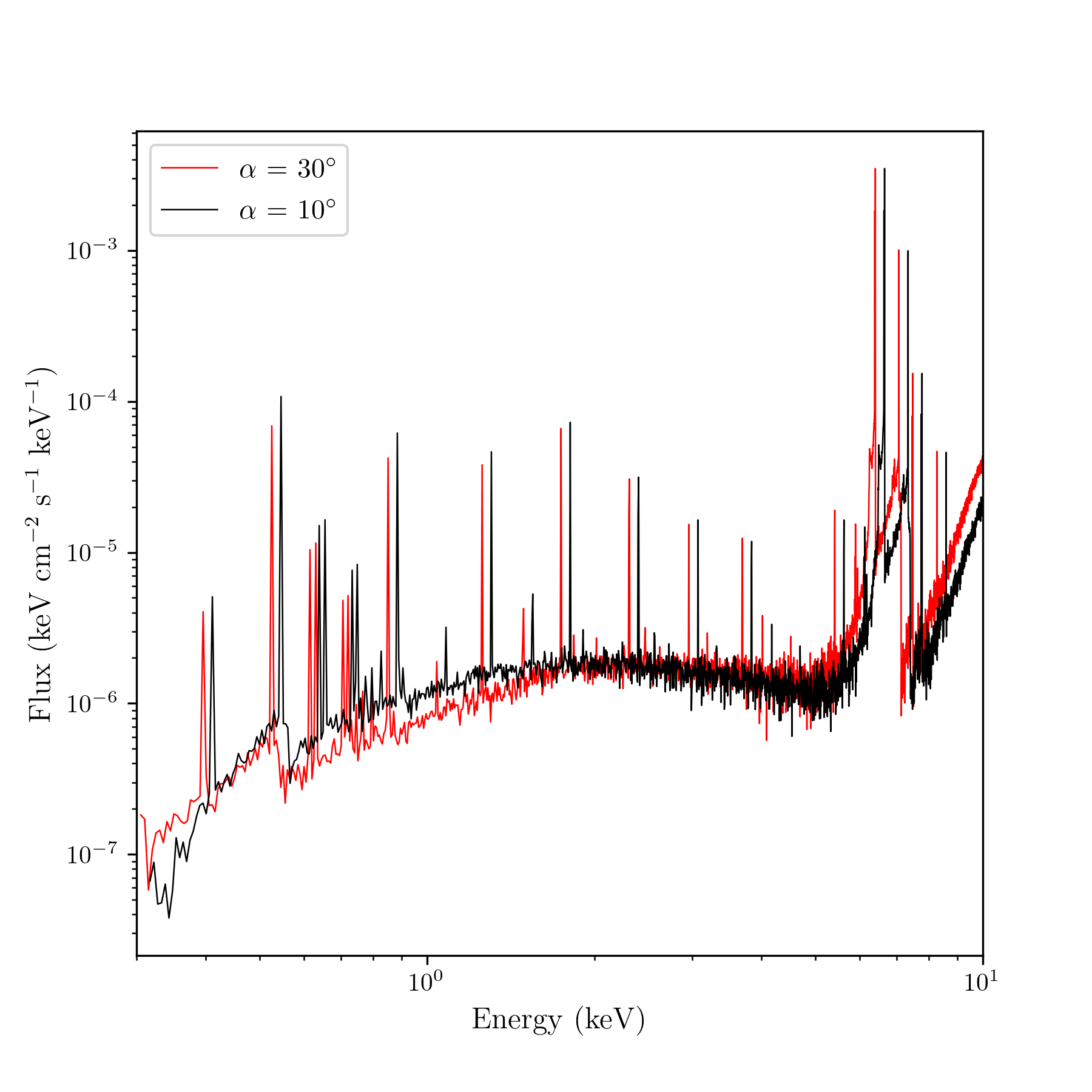}
                \includegraphics[scale = 0.55]{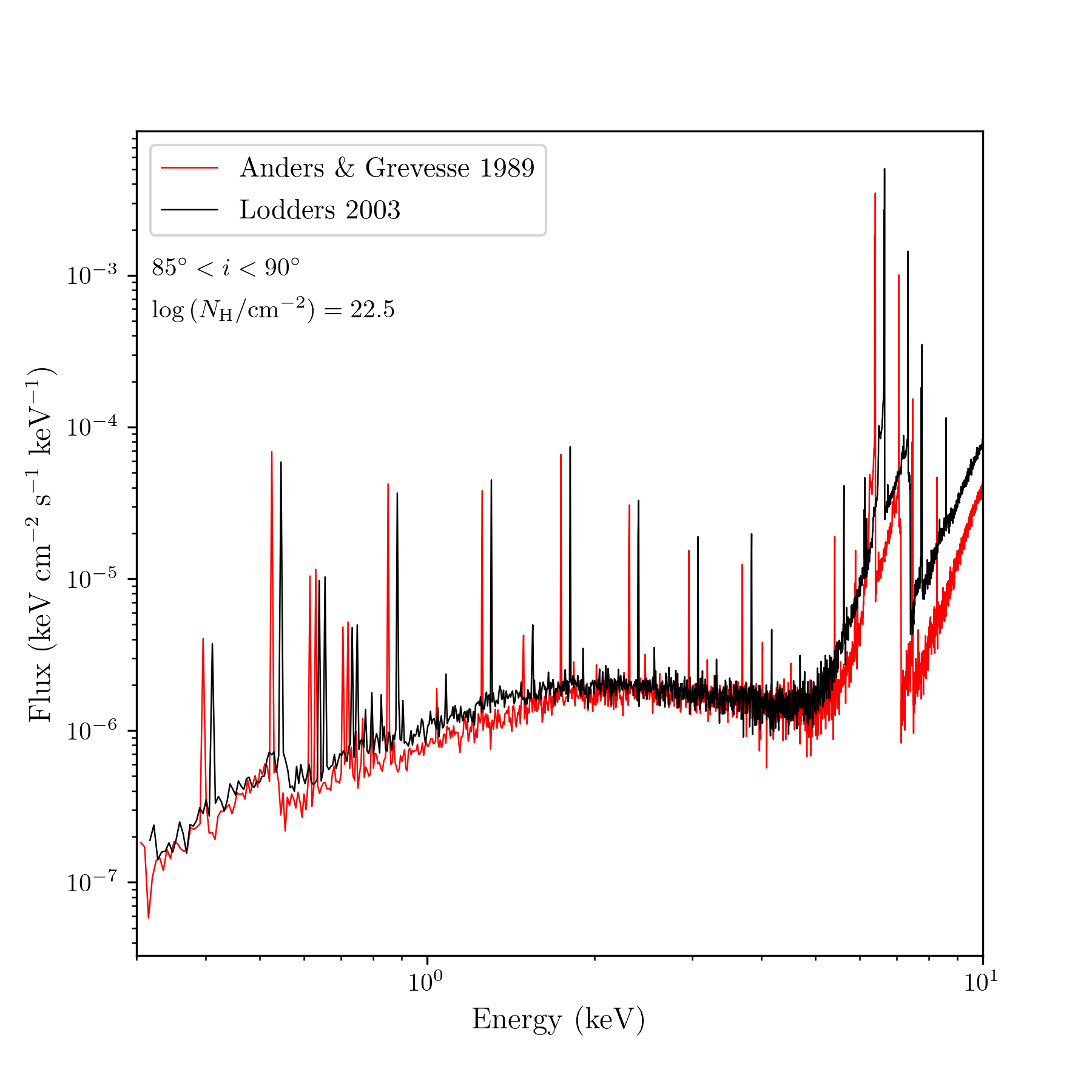}
                \caption{(Left panel) Resulting spectra when considering an opening angle of $\alpha = 10^{\circ}$ for an edge-on torus + hollow cone at the highest polar gas column density (shown in black). The soft spectra (0.3$-$4\,keV) for this new opening angle is best modeled with a power-law with a photon index of $\Gamma = -0.71$ while the $\alpha = 30^{\circ}$ case (shown in red) is best modeled with a photon index of $\Gamma = -0.80$. The EWs for the $\alpha = 10^{\circ}$ case are shown in Table \ref{table7}. (Right panel) Comparison of simulations with a torus + hollow cone at the edge-on inclination with polar gas column density of $N_{\text{H}} = 10^{22.5}$\,cm$^{-2}$ using matter compositions from \citeauthor{1989GeCoA..53..197A} (\citeyear{1989GeCoA..53..197A}; red) and \citeauthor{2003ApJ...591.1220L} (\citeyear{2003ApJ...591.1220L}; black) with H$_{2}$ fractions of 1 and 0.2 respectively. The simulations obtained by considering an opening angle of $\alpha = 30^{\circ}$, as well as a matter composition from \protect\cite{2003ApJ...591.1220L} are shifted to the right by 4$\%$ of the energy in order to clearly see the difference in the spectral lines. There is little spectral difference either in the lines or continuum between the two compositions.}
                \label{fig10}
            \end{figure*}
        
            Looking at the simulations in the soft X-ray band for the torus with a hollow cone gives rise to a question: would an observer be able to distinguish between the edge-on case with a torus + hollow cone at the highest polar gas column density ($N_{\text{H}} = 10^{22.5}$\,cm$^{-2}$) against the intermediate case ($45.6^{\circ} \leq i \leq 50.6^{\circ}$) with a torus alone (blue line in Figure\,\ref{fig2})? Both cases show strong fluorescence lines in the soft X-ray band along with very similar features in the hard X-ray band.
            
            In order to distinguish between the spectra obtained by these two geometries, we fit the slope of the continuum in the soft X-ray band (0.3$-$3\,keV) with simple power-laws. When fitting the spectra, fluorescence lines were not masked when computing the slope, since most instruments cannot mask these lines. The torus alone was best fit with a photon index of $\Gamma = -1.63$. However, the power-law  fit for the edge-on case with a torus + hollow cone gave a photon index of $\Gamma = -0.80$. A plot of the soft spectrum for both simulations considered is shown in Figure\,\ref{fig8.5}. Another approach to distinguish between these two spectra is compare the difference in equivalent widths of the strongest spectral lines in the soft X-ray band for intermediate case with a torus alone and the edge-on case with a torus + hollow cone (the soft spectra for both these cases are shown in Figure\,\ref{fig8.5}). The equivalent widths of the O, Ne, and Si lines were compared for the two geometries. The equivalent widths of these lines are 569, 171, and 173\,eV respectively for the torus alone and 944, 321, and 203\,eV respectively for the torus + hollow cone. All these lines are stronger in the torus + hollow cone scenario with the O line showing the strongest difference, increasing by 375\,eV, compared to the torus-only case.
            
            However, It should be noted that this X-ray emission from the polar gas might be combined with other features, such as those stemming from the NLR on scales of hundreds of parsecs (e,g., \citealp{2006A&A...448..499B, 2007MNRAS.374.1290G}). These contributions include an extra scattered power-law component, as well as the addition of many narrow photoionized lines. The contribution from the NLR to the scattered flux is typically $\sim$0.1--1\% \citep{Ueda_2007,2017ApJS..233...17R, 2021MNRAS.504..428G}, with the lowest values typically associated to the most obscured objects \citep{2021MNRAS.504..428G}. In Figure\,\ref{fig10.5} we add, to the edge-on simulation with a torus + hollow cone at the highest polar gas column density, a power-law component to reproduce a 0.1\% scattered fraction of the intrinsic flux of Circinus, as well as photoionized lines from the obscured AGN NGC 3393 \citep{2006A&A...448..499B} and the Circinus galaxy (\citealp{2001ApJ...546L..13S, 2014ApJ...791...81A}, Andonie et al. submitted), in order to verify how visible the lines in the soft X-ray band will be. The normalisation of these lines were set as to keep the ratio of the flux of line to the power-law the same as in their original source. The value of the extra scattered fraction was selected in order to be consistent with an AGN obscured by CT column densities \citep{2021MNRAS.504..428G}. The resulting spectrum is shown in blue, with the original simulation in dash-dotted black, the scattered power-law as solid black, and the extra photoionized lines dashed and labeled with their transitions. These photoionized lines were chosen from Tables 1, 4, A1 of \citealp{2001ApJ...546L..13S, 2014ApJ...791...81A}, Andonie et al. submitted, as well as Table 4 in \cite{2006A&A...448..499B} based on their proximity to our simulated lines. These new photoionized lines can be distinguished from the O, Ne, and Mg lines. However, the Mg XII - Si II-VI, S II-X, and Ar II-XI transitions have the potential to interfere with the detection of the Si, S, and Ar lines from the polar component. But if the bulk motion of the polar component is blue-shifted enough, these lines could still be distinguished. With this in mind, we calculated the EWs of the O, Ne, Mg and Si (assuming no interference from the photoionized lines) lines which had values of 7\,eV, 12\,eV, 33\,eV, and 87\,eV respectively. 
            
            Additionally, X-ray emission in obscured AGN below $\sim$ 2\,keV caused by populations of X-ray binaries and collisionally ionised plazma in star forming regions (e.g., \citealp{2008MNRAS.386.1464R}) could also pose a challenge in detecting the true emission from the polar component. To consider this additional contribution, we added to the model shown in Figure\,\ref{fig10.5} a thermal plasma model with a temperature of 0.5\,keV which corresponds to the median temperature in nearby obscured AGN \citep{2017ApJS..233...17R}. We found the addition of this component should not add additional difficulty in detecting the spectral lines from the polar component in obscured AGN.
            
        \subsection{Observations with \textit{XRISM} and \textit{Athena}}
        
            Here we consider how our previous simulations of the polar component would be observed with instruments on board \textit{XRISM} (Resolve; \citealp{xrismscienceteam2020science}) and \textit{Athena} (X-IFU; \citealp{2016SPIE.9905E..2FB}). \textit{XRISM}/Resolve boasts a $\sim 5-7$\,eV FWHM spectral resolution over the entire bandpass, while \textit{Athena}/X-IFU in expected to operate under a $\sim 2.5$\,eV spectral resolution up to 7\,keV. Based on results of our spectral simulations, these instruments could be extremely well suited to identify the X-ray signature of the polar component. We use the X-ray spectral-fitting program software \textsc{XSPEC} \citep{1996ASPC..101...17A} along with the latest \textit{XRISM}/Resolve and \textit{Athena}/X-IFU (Barret et al. 2021 in prep.) response and background files to simulate the spectra that would be expected from these two instruments. The spectral simulations are illustrated in Figure\,\ref{figres}, where we consider the edge-on torus + hollow cone geometry, assuming the highest polar gas column density ($\log{N_{\mathrm{H}}\big/\text{cm}^{-2}} = 22.5$; left panel of Figure\,\ref{fig3b}), along with the components discussed in section 4.2 (power-law with 0.1\% of the scattered flux with various photoionized lines from the NLR). We assumed an exposure time of 200\,ks with an observed flux in the $2-10$\,keV range of $F_{2-10} = 12.7\times10^{-12}$\,erg s$^{-1}$ cm$^{-2}$, as expected for the Circinus galaxy \citep{2017ApJS..233...17R}. The top panel displays the \textit{XRISM}/Resolve response while the bottom shows the \textit{Athena}/X-IFU response. Based on this figure, \textit{Athena}/X-IFU will be better suited to detect lines from the polar component, since \textit{XRISM}/Resolve is unable to distinguish between some of the lines from the polar component and and photoionized lines from the NLR. However, \textit{XRISM}/Resolve is still able to resolve the Ca line at $\sim$ 3.69\,keV. \textit{Athena}/X-IFU is able to clearly distinguish the O, Ne, Mg, and Ca lines from the NLR lines. It should be stressed that, if this polar component in caused by out-flowing gas \citep{Ricci2017, Leftley_2019, 2019ApJ...884..171H, 2020ApJ...900..174V}, it is likely that these lines will be shifted to different energies. We expect, given the resolutions of \textit{XRISM}/Resolve and \textit{Athena}/X-IFU, to be able to recover velocity shifts of $\sim$ 1000\,km/s and 375\,km/s for the two instruments respectively for the O, F, Ne, Mg, and Si lines.     
            
    \section{Testing Different Cone Geometries and Abundances}
    
        \subsection{Changing the Opening Angle}
    
            It is possible that the opening angle of the polar component could be controlled through collimation by the absorbing material surrounding the BLR and acccretion disk (i.e., the torus), as is the case for the NLR \citep{2013ApJS..209....1F}. With this in mind, we considered a new opening angle of $\alpha = 10^{\circ}$ to simulate a nucleus with higher collimation of ionizing radiation which corresponds to a torus with a larger covering factor. The resulting spectra compared with the previous opening angle of $\alpha = 30^{\circ}$ for an edge-on torus + hollow cone at the highest column density can be seen on the left panel of in Figure\,\ref{fig10}. Fitting the soft spectra with a simple power-law for the $\alpha = 10^{\circ}$ case gives a photon index of $\Gamma = -0.71$ compared to an index of $\Gamma = -0.80$ for the $\alpha = 30^{\circ}$ case. Table \ref{table7} shows all the EWs of the lines in the soft spectra for this new opening angle for different densities of the polar gas as well. We note that, while the O line has a higher EW for the  $\alpha = 30^{\circ}$ case (an average ratio of 1.29 when averaging with all four column densities), the Ne and Si lines tend to have larger EWs in the  $\alpha = 10^{\circ}$ case with average ratios of 0.79 and 0.99 respectively.
            
        \subsection{Changing Abundances}
        
            The simulations carried out so far considered a matter composition from \cite{1989GeCoA..53..197A} as well as a molecular hydrogen fraction of one. In this section, we consider the more recent gas composition from \cite{2003ApJ...591.1220L} as well as a Hydrogen fraction of 0.2 (H$_{2} = 0.2$). We ran simulations using this new matter composition for the edge-on case at the highest polar gas column density in order to compare with the previous simulations. The results are seen on the right panel of Figure\,\ref{fig10} with the black plot representing the composition from \cite{2003ApJ...591.1220L} (H$_{2}$ = 0.2) and the red plot being from the previous simulations using \cite{1989GeCoA..53..197A} (H$_{2}$ = 1). As can be seen, there is no significant difference in spectral or continuum features. 
    
    \section{Summary and Conclusion}
    
        In this paper we have investigated the effect polar gas has on the X-ray spectra of accreting supermassive black holes. Three geometries (torus alone, torus + hollow cone, torus + filled cone), along with three inclination angles and four slant length column densities for the filled and hollow cones were considered. We summarize our main results below:
        
        \begin{itemize}
            \item The polar gas was found to have a significant impact on the soft X-ray spectrum producing several fluorescence lines in the 0.3--5\,keV band such as the O ($524.9$\,eV), Ne ($848.6$\,eV), Mg ($1253.6$\,eV), and Si ($1739.38$\,eV) lines. The most significant impact on the X-ray spectra from the polar gas occurs at the edge-on inclination angle where photons which would normally be absorbed in the torus are allowed to scatter from the less dense polar component (see Figures\,\ref{fig2}, \ref{fig3a} and \ref{fig3b}, as well as Figure\,\ref{fig6} and Tables \ref{table1}-\ref{table6} for the EWs of selected spectral lines). 
            \item The most significant impact the polar component has is when it is in the form of a hollow cone, as the filled cone causes more self-absorption as there is more gas to interact with. Simulations were also run using the same number of atoms in the filled as in the hollow cone. These showed that the soft spectrum of the pole-on case was still obscured and, therefore, is likely not the geometry for the polar component (see Figure \,\ref{figCone}). 
            \item When considering 0.1\% of extra scattered continuum stemming from the NLR in addition to the new polar gas emission, as well as extra photoionized lines from NGC 3393 and the Circinus galaxy, we find that the spectral lines from the polar component would still be detectable, particularly by future X-ray calorimeters such as those onboard \textit{XRISM} (Resolve; \citealp{xrismscienceteam2020science}) and \textit{Athena} (X-IFU; \citealp{2016SPIE.9905E..2FB}) (see Figure\,\ref{fig10.5}). We also simulate spectra using \textit{XRISM} and \textit{Athena} (Barret et al. 2021 in prep.) to see how our simulated data would be observed (see Figure\,\ref{figres}). This figure shows that \textit{Athena}/X-IFU is better suited to observe more spectral features from the polar component than \textit{XRISM}/Resolve, given the extra features stemming from the NLR.  
            \item When considering a smaller opening angle for the polar component we find that, while the EWs of many spectral lines change, the general trends as seen in Figure\,\ref{fig6} remain the same with the EWs of lower energy lines (O, F, Ne, Mg, Si) increasing with increasing column density of the polar component and the EWs of the higher energy lines (Ar, Ca) decreasing with column density. This is with the exception of the S line which increases in EW, contrary to what is seen in Figure\,\ref{fig6} (see Table \ref{table7}). In addition to this, when considering an abundance from \cite{2003ApJ...591.1220L}, we find little difference in spectral or continuum features (see the right panel of Figure\,\ref{fig10}).  
        \end{itemize}
        
        These simulations show that low-energy fluorescent lines could be an important tracer of the polar gas in AGN. Observations of nearby AGN with future high-resolution X-ray instruments, such as those on-board {\it XRISM} \citep{xrismscienceteam2020science} and {\it Athena} X-IFU \citep{2016SPIE.9905E..2FB}, will allow to search for signatures of polar gas. Comparison with simulations, such as those reported in this paper, will allow further constrain the kinematics, geometry, and origins of the polar gas.




\newpage




\bibliographystyle{mnras}
\bibliography{bibi} 

\begin{thebibliography}{}
\makeatletter
\relax
\def\mn@urlcharsother{\let\do\@makeother \do\$\do\&\do\#\do\^\do\_\do\%\do\~}
\def\mn@doi{\begingroup\mn@urlcharsother \@ifnextchar [ {\mn@doi@}
  {\mn@doi@[]}}
\def\mn@doi@[#1]#2{\def\@tempa{#1}\ifx\@tempa\@empty \href
  {http://dx.doi.org/#2} {doi:#2}\else \href {http://dx.doi.org/#2} {#1}\fi
  \endgroup}
\def\mn@eprint#1#2{\mn@eprint@#1:#2::\@nil}
\def\mn@eprint@arXiv#1{\href {http://arxiv.org/abs/#1} {{\tt arXiv:#1}}}
\def\mn@eprint@dblp#1{\href {http://dblp.uni-trier.de/rec/bibtex/#1.xml}
  {dblp:#1}}
\def\mn@eprint@#1:#2:#3:#4\@nil{\def\@tempa {#1}\def\@tempb {#2}\def\@tempc
  {#3}\ifx \@tempc \@empty \let \@tempc \@tempb \let \@tempb \@tempa \fi \ifx
  \@tempb \@empty \def\@tempb {arXiv}\fi \@ifundefined
  {mn@eprint@\@tempb}{\@tempb:\@tempc}{\expandafter \expandafter \csname
  mn@eprint@\@tempb\endcsname \expandafter{\@tempc}}}

\bibitem[\protect\citeauthoryear{{Anders} \& {Grevesse}}{{Anders} \&
  {Grevesse}}{1989}]{1989GeCoA..53..197A}
{Anders} E.,  {Grevesse} N.,  1989, \mn@doi [\gca]
  {10.1016/0016-7037(89)90286-X}, \href
  {https://ui.adsabs.harvard.edu/abs/1989GeCoA..53..197A} {53, 197}

\bibitem[\protect\citeauthoryear{{Antonucci}}{{Antonucci}}{1993}]{1993ARA&A..31..473A}
{Antonucci} R.,  1993, \mn@doi [\araa] {10.1146/annurev.aa.31.090193.002353},
  \href {https://ui.adsabs.harvard.edu/abs/1993ARA&A..31..473A} {31, 473}

\bibitem[\protect\citeauthoryear{{Ar{\'e}valo} et~al.,}{{Ar{\'e}valo}
  et~al.}{2014}]{2014ApJ...791...81A}
{Ar{\'e}valo} P.,  et~al., 2014, \mn@doi [\apj] {10.1088/0004-637X/791/2/81},
  \href {https://ui.adsabs.harvard.edu/abs/2014ApJ...791...81A} {791, 81}

\bibitem[\protect\citeauthoryear{{Arnaud}}{{Arnaud}}{1996}]{1996ASPC..101...17A}
{Arnaud} K.~A.,  1996, in {Jacoby} G.~H.,  {Barnes} J.,  eds,  Astronomical
  Society of the Pacific Conference Series Vol. 101, Astronomical Data Analysis
  Software and Systems V. p.~17

\bibitem[\protect\citeauthoryear{Asmus}{Asmus}{2019}]{10.1093/mnras/stz2289}
Asmus D.,  2019, \mn@doi [Monthly Notices of the Royal Astronomical Society]
  {10.1093/mnras/stz2289}, 489, 2177

\bibitem[\protect\citeauthoryear{Asmus, Hönig  \& Gandhi}{Asmus
  et~al.}{2016}]{Asmus_2016}
Asmus D.,  Hönig S.~F.,   Gandhi P.,  2016, \mn@doi [The Astrophysical
  Journal] {10.3847/0004-637x/822/2/109}, 822, 109

\bibitem[\protect\citeauthoryear{{Balokovi{\'c}} et~al.,}{{Balokovi{\'c}}
  et~al.}{2018}]{2018ApJ...854...42B}
{Balokovi{\'c}} M.,  et~al., 2018, \mn@doi [\apj] {10.3847/1538-4357/aaa7eb},
  \href {https://ui.adsabs.harvard.edu/abs/2018ApJ...854...42B} {854, 42}

\bibitem[\protect\citeauthoryear{{Barret} et~al.,}{{Barret}
  et~al.}{2016}]{2016SPIE.9905E..2FB}
{Barret} D.,  et~al., 2016, in {den Herder} J.-W.~A.,  {Takahashi} T.,
  {Bautz} M.,  eds,  Society of Photo-Optical Instrumentation Engineers (SPIE)
  Conference Series Vol. 9905, Space Telescopes and Instrumentation 2016:
  Ultraviolet to Gamma Ray. p. 99052F (\mn@eprint {arXiv} {1608.08105}),
  \mn@doi{10.1117/12.2232432}

\bibitem[\protect\citeauthoryear{{Bianchi}, {Guainazzi}  \&
  {Chiaberge}}{{Bianchi} et~al.}{2006}]{2006A&A...448..499B}
{Bianchi} S.,  {Guainazzi} M.,   {Chiaberge} M.,  2006, \mn@doi [\aap]
  {10.1051/0004-6361:20054091}, \href
  {https://ui.adsabs.harvard.edu/abs/2006A&A...448..499B} {448, 499}

\bibitem[\protect\citeauthoryear{{Brightman} \& {Nandra}}{{Brightman} \&
  {Nandra}}{2011b}]{2011MNRAS.413.1206B}
{Brightman} M.,  {Nandra} K.,  2011b, \mn@doi [\mnras]
  {10.1111/j.1365-2966.2011.18207.x}, \href
  {https://ui.adsabs.harvard.edu/abs/2011MNRAS.413.1206B} {413, 1206}

\bibitem[\protect\citeauthoryear{Brightman \& Nandra}{Brightman \&
  Nandra}{2011a}]{10.1111/j.1365-2966.2011.18207.x}
Brightman M.,  Nandra K.,  2011a, \mn@doi [Monthly Notices of the Royal
  Astronomical Society] {10.1111/j.1365-2966.2011.18207.x}, 413, 1206

\bibitem[\protect\citeauthoryear{Brightman \& Ueda}{Brightman \&
  Ueda}{2012}]{10.1111/j.1365-2966.2012.20908.x}
Brightman M.,  Ueda Y.,  2012, \mn@doi [Monthly Notices of the Royal
  Astronomical Society] {10.1111/j.1365-2966.2012.20908.x}, 423, 702

\bibitem[\protect\citeauthoryear{{Buchner}, {Brightman}, {Nandra}, {Nikutta}
  \& {Bauer}}{{Buchner} et~al.}{2019}]{2019A&A...629A..16B}
{Buchner} J.,  {Brightman} M.,  {Nandra} K.,  {Nikutta} R.,   {Bauer} F.~E.,
  2019, \mn@doi [\aap] {10.1051/0004-6361/201834771}, \href
  {https://ui.adsabs.harvard.edu/abs/2019A&A...629A..16B} {629, A16}

\bibitem[\protect\citeauthoryear{{Burtscher} et~al.,}{{Burtscher}
  et~al.}{2013}]{2013A&A...558A.149B}
{Burtscher} L.,  et~al., 2013, \mn@doi [\aap] {10.1051/0004-6361/201321890},
  \href {https://ui.adsabs.harvard.edu/abs/2013A&A...558A.149B} {558, A149}

\bibitem[\protect\citeauthoryear{Burtscher, Hönig, Jaffe, Kishimoto,
  Lopez-Gonzaga, Meisenheimer  \& Tristam}{Burtscher
  et~al.}{2016}]{10.1117/12.2231077}
Burtscher L.,  Hönig S.,  Jaffe W.,  Kishimoto M.,  Lopez-Gonzaga N.,
  Meisenheimer K.,   Tristam K. R.~W.,  2016, in Malbet F.,  Creech-Eakman
  M.~J.,   Tuthill P.~G.,  eds,  Vol. 9907, Optical and Infrared Interferometry
  and Imaging V. SPIE, pp 173 -- 186, \mn@doi{10.1117/12.2231077}, \url
  {https://doi.org/10.1117/12.2231077}

\bibitem[\protect\citeauthoryear{{Chartas}, {Kochanek}, {Dai}, {Poindexter}  \&
  {Garmire}}{{Chartas} et~al.}{2009}]{2009ApJ...693..174C}
{Chartas} G.,  {Kochanek} C.~S.,  {Dai} X.,  {Poindexter} S.,   {Garmire} G.,
  2009, \mn@doi [\apj] {10.1088/0004-637X/693/1/174}, \href
  {https://ui.adsabs.harvard.edu/abs/2009ApJ...693..174C} {693, 174}

\bibitem[\protect\citeauthoryear{{Fischer}, {Crenshaw}, {Kraemer}  \&
  {Schmitt}}{{Fischer} et~al.}{2013}]{2013ApJS..209....1F}
{Fischer} T.~C.,  {Crenshaw} D.~M.,  {Kraemer} S.~B.,   {Schmitt} H.~R.,  2013,
  \mn@doi [\apjs] {10.1088/0067-0049/209/1/1}, \href
  {https://ui.adsabs.harvard.edu/abs/2013ApJS..209....1F} {209, 1}

\bibitem[\protect\citeauthoryear{{Gallimore} et~al.,}{{Gallimore}
  et~al.}{2016}]{2016ApJ...829L...7G}
{Gallimore} J.~F.,  et~al., 2016, \mn@doi [\apjl] {10.3847/2041-8205/829/1/L7},
  \href {https://ui.adsabs.harvard.edu/abs/2016ApJ...829L...7G} {829, L7}

\bibitem[\protect\citeauthoryear{{George} \& {Fabian}}{{George} \&
  {Fabian}}{1991}]{1991MNRAS.249..352G}
{George} I.~M.,  {Fabian} A.~C.,  1991, \mn@doi [\mnras]
  {10.1093/mnras/249.2.352}, \href
  {https://ui.adsabs.harvard.edu/abs/1991MNRAS.249..352G} {249, 352}

\bibitem[\protect\citeauthoryear{{Guainazzi} \& {Bianchi}}{{Guainazzi} \&
  {Bianchi}}{2007}]{2007MNRAS.374.1290G}
{Guainazzi} M.,  {Bianchi} S.,  2007, \mn@doi [\mnras]
  {10.1111/j.1365-2966.2006.11229.x}, \href
  {https://ui.adsabs.harvard.edu/abs/2007MNRAS.374.1290G} {374, 1290}

\bibitem[\protect\citeauthoryear{{Guilbert} \& {Rees}}{{Guilbert} \&
  {Rees}}{1988}]{1988MNRAS.233..475G}
{Guilbert} P.~W.,  {Rees} M.~J.,  1988, \mn@doi [\mnras]
  {10.1093/mnras/233.2.475}, \href
  {https://ui.adsabs.harvard.edu/abs/1988MNRAS.233..475G} {233, 475}

\bibitem[\protect\citeauthoryear{{Gupta} et~al.,}{{Gupta}
  et~al.}{2021}]{2021MNRAS.504..428G}
{Gupta} K.~K.,  et~al., 2021, \mn@doi [\mnras] {10.1093/mnras/stab839}, \href
  {https://ui.adsabs.harvard.edu/abs/2021MNRAS.504..428G} {504, 428}

\bibitem[\protect\citeauthoryear{{Haardt} \& {Maraschi}}{{Haardt} \&
  {Maraschi}}{1991}]{1991ApJ...380L..51H}
{Haardt} F.,  {Maraschi} L.,  1991, \mn@doi [\apjl] {10.1086/186171}, \href
  {https://ui.adsabs.harvard.edu/abs/1991ApJ...380L..51H} {380, L51}

\bibitem[\protect\citeauthoryear{{Haardt} \& {Maraschi}}{{Haardt} \&
  {Maraschi}}{1993}]{1993ApJ...413..507H}
{Haardt} F.,  {Maraschi} L.,  1993, \mn@doi [\apj] {10.1086/173020}, \href
  {https://ui.adsabs.harvard.edu/abs/1993ApJ...413..507H} {413, 507}

\bibitem[\protect\citeauthoryear{{Haardt}, {Maraschi}  \&
  {Ghisellini}}{{Haardt} et~al.}{1994}]{1994ApJ...432L..95H}
{Haardt} F.,  {Maraschi} L.,   {Ghisellini} G.,  1994, \mn@doi [\apjl]
  {10.1086/187520}, \href
  {https://ui.adsabs.harvard.edu/abs/1994ApJ...432L..95H} {432, L95}

\bibitem[\protect\citeauthoryear{{Henri} \& {Petrucci}}{{Henri} \&
  {Petrucci}}{1997}]{1997A&A...326...87H}
{Henri} G.,  {Petrucci} P.~O.,  1997, \aap, \href
  {https://ui.adsabs.harvard.edu/abs/1997A&A...326...87H} {326, 87}

\bibitem[\protect\citeauthoryear{Hikitani, Ohno, Fukazawa, Kawaguchi  \&
  Odaka}{Hikitani et~al.}{2018}]{Hikitani_2018}
Hikitani M.,  Ohno M.,  Fukazawa Y.,  Kawaguchi T.,   Odaka H.,  2018, \mn@doi
  [The Astrophysical Journal] {10.3847/1538-4357/aae1fe}, 867, 80

\bibitem[\protect\citeauthoryear{{H{\"o}nig}}{{H{\"o}nig}}{2019}]{2019ApJ...884..171H}
{H{\"o}nig} S.~F.,  2019, \mn@doi [\apj] {10.3847/1538-4357/ab4591}, \href
  {https://ui.adsabs.harvard.edu/abs/2019ApJ...884..171H} {884, 171}

\bibitem[\protect\citeauthoryear{{H{\"o}nig}, {Kishimoto}, {Antonucci},
  {Marconi}, {Prieto}, {Tristram}  \& {Weigelt}}{{H{\"o}nig}
  et~al.}{2012}]{2012ApJ...755..149H}
{H{\"o}nig} S.~F.,  {Kishimoto} M.,  {Antonucci} R.,  {Marconi} A.,  {Prieto}
  M.~A.,  {Tristram} K.,   {Weigelt} G.,  2012, \mn@doi [\apj]
  {10.1088/0004-637X/755/2/149}, \href
  {https://ui.adsabs.harvard.edu/abs/2012ApJ...755..149H} {755, 149}

\bibitem[\protect\citeauthoryear{Hönig, Kishimoto, Antonucci, Marconi, Prieto,
  Tristram  \& Weigelt}{Hönig et~al.}{2012}]{H_nig_2012}
Hönig S.~F.,  Kishimoto M.,  Antonucci R.,  Marconi A.,  Prieto M.~A.,
  Tristram K.,   Weigelt G.,  2012, \mn@doi [The Astrophysical Journal]
  {10.1088/0004-637x/755/2/149}, 755, 149

\bibitem[\protect\citeauthoryear{Jaffe et~al.,}{Jaffe et~al.}{2004}]{Jaffe2004}
Jaffe W.,  et~al., 2004, \mn@doi [Nature] {10.1038/nature02531}, 429, 47

\bibitem[\protect\citeauthoryear{{Kishimoto}, {H{\"o}nig}, {Beckert}  \&
  {Weigelt}}{{Kishimoto} et~al.}{2007}]{2007A&A...476..713K}
{Kishimoto} M.,  {H{\"o}nig} S.~F.,  {Beckert} T.,   {Weigelt} G.,  2007,
  \mn@doi [\aap] {10.1051/0004-6361:20077911}, \href
  {https://ui.adsabs.harvard.edu/abs/2007A&A...476..713K} {476, 713}

\bibitem[\protect\citeauthoryear{{Koss} et~al.,}{{Koss}
  et~al.}{2017}]{2017ApJ...850...74K}
{Koss} M.,  et~al., 2017, \mn@doi [\apj] {10.3847/1538-4357/aa8ec9}, \href
  {https://ui.adsabs.harvard.edu/abs/2017ApJ...850...74K} {850, 74}

\bibitem[\protect\citeauthoryear{Leftley, Hönig, Asmus, Tristram, Gandhi,
  Kishimoto, Venanzi  \& Williamson}{Leftley et~al.}{2019}]{Leftley_2019}
Leftley J.~H.,  Hönig S.~F.,  Asmus D.,  Tristram K. R.~W.,  Gandhi P.,
  Kishimoto M.,  Venanzi M.,   Williamson D.~J.,  2019, \mn@doi [The
  Astrophysical Journal] {10.3847/1538-4357/ab4a0b}, 886, 55

\bibitem[\protect\citeauthoryear{{Liu} \& {Li}}{{Liu} \&
  {Li}}{2014}]{2014ApJ...787...52L}
{Liu} Y.,  {Li} X.,  2014, \mn@doi [\apj] {10.1088/0004-637X/787/1/52}, \href
  {https://ui.adsabs.harvard.edu/abs/2014ApJ...787...52L} {787, 52}

\bibitem[\protect\citeauthoryear{{Liu}, {H{\"o}nig}, {Ricci}  \&
  {Paltani}}{{Liu} et~al.}{2019}]{Liu2019XraySO}
{Liu} J.,  {H{\"o}nig} S.~F.,  {Ricci} C.,   {Paltani} S.,  2019, \mn@doi
  [\mnras] {10.1093/mnras/stz2908}, \href
  {https://ui.adsabs.harvard.edu/abs/2019MNRAS.490.4344L} {490, 4344}

\bibitem[\protect\citeauthoryear{{Lodders}}{{Lodders}}{2003}]{2003ApJ...591.1220L}
{Lodders} K.,  2003, \mn@doi [\apj] {10.1086/375492}, \href
  {https://ui.adsabs.harvard.edu/abs/2003ApJ...591.1220L} {591, 1220}

\bibitem[\protect\citeauthoryear{López~Gonzaga, Jaffe, Burtscher, Tristram  \&
  Meisenheimer}{López~Gonzaga et~al.}{2014}]{refId01}
López~Gonzaga N.,  Jaffe W.,  Burtscher L.,  Tristram K.,   Meisenheimer K.,
  2014, \mn@doi [Astronomy & Astrophysics] {10.1051/0004-6361/201323002}, 565

\bibitem[\protect\citeauthoryear{López~Gonzaga, Burtscher, Tristram,
  Meisenheimer  \& Schartmann}{López~Gonzaga et~al.}{2016}]{refId03}
López~Gonzaga N.,  Burtscher L.,  Tristram K.,  Meisenheimer K.,   Schartmann
  M.,  2016, \mn@doi [Astronomy & Astrophysics] {10.1051/0004-6361/201527590},
  591

\bibitem[\protect\citeauthoryear{{Magorrian} et~al.,}{{Magorrian}
  et~al.}{1998}]{1998AJ....115.2285M}
{Magorrian} J.,  et~al., 1998, \mn@doi [\aj] {10.1086/300353}, \href
  {https://ui.adsabs.harvard.edu/abs/1998AJ....115.2285M} {115, 2285}

\bibitem[\protect\citeauthoryear{{Matt}}{{Matt}}{2002}]{2002MNRAS.337..147M}
{Matt} G.,  2002, \mn@doi [\mnras] {10.1046/j.1365-8711.2002.05890.x}, \href
  {https://ui.adsabs.harvard.edu/abs/2002MNRAS.337..147M} {337, 147}

\bibitem[\protect\citeauthoryear{Matt, Fabian, Guainazzi, Iwasawa, Bassani  \&
  Malaguti}{Matt et~al.}{2000}]{10.1046/j.1365-8711.2000.03721.x}
Matt G.,  Fabian A.~C.,  Guainazzi M.,  Iwasawa K.,  Bassani L.,   Malaguti G.,
   2000, \mn@doi [Monthly Notices of the Royal Astronomical Society]
  {10.1046/j.1365-8711.2000.03721.x}, 318, 173

\bibitem[\protect\citeauthoryear{{Merloni}, {Di Matteo}  \& {Fabian}}{{Merloni}
  et~al.}{2000}]{2000MNRAS.318L..15M}
{Merloni} A.,  {Di Matteo} T.,   {Fabian} A.~C.,  2000, \mn@doi [\mnras]
  {10.1046/j.1365-8711.2000.03943.x}, \href
  {https://ui.adsabs.harvard.edu/abs/2000MNRAS.318L..15M} {318, L15}

\bibitem[\protect\citeauthoryear{Murphy \& Yaqoob}{Murphy \&
  Yaqoob}{2009}]{10.1111/j.1365-2966.2009.15025.x}
Murphy K.~D.,  Yaqoob T.,  2009, \mn@doi [Monthly Notices of the Royal
  Astronomical Society] {10.1111/j.1365-2966.2009.15025.x}, 397, 1549

\bibitem[\protect\citeauthoryear{{Nandra}, {Le}, {George}, {Edelson},
  {Mushotzky}, {Peterson}  \& {Turner}}{{Nandra} et~al.}{2000}]{Nandra7469}
{Nandra} K.,  {Le} T.,  {George} I.~M.,  {Edelson} R.~A.,  {Mushotzky} R.~F.,
  {Peterson} B.~M.,   {Turner} T.~J.,  2000, \apj, 544

\bibitem[\protect\citeauthoryear{{Nenkova}, {Sirocky}, {Ivezi{\'c}}  \&
  {Elitzur}}{{Nenkova} et~al.}{2008}]{2008ApJ...685..147N}
{Nenkova} M.,  {Sirocky} M.~M.,  {Ivezi{\'c}} {\v{Z}}.,   {Elitzur} M.,  2008,
  \mn@doi [\apj] {10.1086/590482}, \href
  {https://ui.adsabs.harvard.edu/abs/2008ApJ...685..147N} {685, 147}

\bibitem[\protect\citeauthoryear{{Netzer}}{{Netzer}}{2015}]{2015ARA&A..53..365N}
{Netzer} H.,  2015, \mn@doi [\araa] {10.1146/annurev-astro-082214-122302},
  \href {https://ui.adsabs.harvard.edu/abs/2015ARA&A..53..365N} {53, 365}

\bibitem[\protect\citeauthoryear{Odaka, Yoneda, Takahashi  \& Fabian}{Odaka
  et~al.}{2016}]{10.1093/mnras/stw1764}
Odaka H.,  Yoneda H.,  Takahashi T.,   Fabian A.,  2016, \mn@doi [Monthly
  Notices of the Royal Astronomical Society] {10.1093/mnras/stw1764}, 462, 2366

\bibitem[\protect\citeauthoryear{Paltani \& Ricci}{Paltani \&
  Ricci}{2017}]{RefleX}
Paltani S.,  Ricci 2017, A\&A, 607, A31

\bibitem[\protect\citeauthoryear{{Ramos Almeida} \& {Ricci}}{{Ramos Almeida} \&
  {Ricci}}{2017}]{2017NatAs...1..679R}
{Ramos Almeida} C.,  {Ricci} C.,  2017, \mn@doi [Nature Astronomy]
  {10.1038/s41550-017-0232-z}, \href
  {https://ui.adsabs.harvard.edu/abs/2017NatAs...1..679R} {1, 679}

\bibitem[\protect\citeauthoryear{{Ranalli}, {Comastri}, {Origlia}  \&
  {Maiolino}}{{Ranalli} et~al.}{2008}]{2008MNRAS.386.1464R}
{Ranalli} P.,  {Comastri} A.,  {Origlia} L.,   {Maiolino} R.,  2008, \mn@doi
  [\mnras] {10.1111/j.1365-2966.2008.13128.x}, \href
  {https://ui.adsabs.harvard.edu/abs/2008MNRAS.386.1464R} {386, 1464}

\bibitem[\protect\citeauthoryear{{Ricci}, {Ueda}, {Koss}, {Trakhtenbrot},
  {Bauer}  \& {Gandhi}}{{Ricci} et~al.}{2015}]{2015ApJ...815L..13R}
{Ricci} C.,  {Ueda} Y.,  {Koss} M.~J.,  {Trakhtenbrot} B.,  {Bauer} F.~E.,
  {Gandhi} P.,  2015, \mn@doi [\apjl] {10.1088/2041-8205/815/1/L13}, \href
  {https://ui.adsabs.harvard.edu/abs/2015ApJ...815L..13R} {815, L13}

\bibitem[\protect\citeauthoryear{{Ricci} et~al.,}{{Ricci}
  et~al.}{2017a}]{2017ApJS..233...17R}
{Ricci} C.,  et~al., 2017a, \mn@doi [\apjs] {10.3847/1538-4365/aa96ad}, \href
  {https://ui.adsabs.harvard.edu/abs/2017ApJS..233...17R} {233, 17}

\bibitem[\protect\citeauthoryear{Ricci et~al.,}{Ricci
  et~al.}{2017b}]{Ricci2017}
Ricci C.,  et~al., 2017b, \mn@doi [Nature] {10.1038/nature23906}, 549, 488

\bibitem[\protect\citeauthoryear{Ricci et~al.,}{Ricci
  et~al.}{2018}]{10.1093/mnras/sty1879}
Ricci C.,  et~al., 2018, \mn@doi [Monthly Notices of the Royal Astronomical
  Society] {10.1093/mnras/sty1879}, 480, 1819

\bibitem[\protect\citeauthoryear{{Sambruna}, {Netzer}, {Kaspi}, {Brandt},
  {Chartas}, {Garmire}, {Nousek}  \& {Weaver}}{{Sambruna}
  et~al.}{2001}]{2001ApJ...546L..13S}
{Sambruna} R.~M.,  {Netzer} H.,  {Kaspi} S.,  {Brandt} W.~N.,  {Chartas} G.,
  {Garmire} G.~P.,  {Nousek} J.~A.,   {Weaver} K.~A.,  2001, \mn@doi [\apjl]
  {10.1086/318068}, \href
  {https://ui.adsabs.harvard.edu/abs/2001ApJ...546L..13S} {546, L13}

\bibitem[\protect\citeauthoryear{{Shakura} \& {Sunyaev}}{{Shakura} \&
  {Sunyaev}}{1973}]{1973A&A....24..337S}
{Shakura} N.~I.,  {Sunyaev} R.~A.,  1973, \aap, \href
  {https://ui.adsabs.harvard.edu/abs/1973A&A....24..337S} {500, 33}

\bibitem[\protect\citeauthoryear{{Stalevski}, {Fritz}, {Baes}, {Nakos}  \&
  {Popovi{\'c}}}{{Stalevski} et~al.}{2012}]{2012MNRAS.420.2756S}
{Stalevski} M.,  {Fritz} J.,  {Baes} M.,  {Nakos} T.,   {Popovi{\'c}}
  L.~{\v{C}}.,  2012, \mn@doi [\mnras] {10.1111/j.1365-2966.2011.19775.x},
  \href {https://ui.adsabs.harvard.edu/abs/2012MNRAS.420.2756S} {420, 2756}

\bibitem[\protect\citeauthoryear{Stalevski, Asmus  \& Tristram}{Stalevski
  et~al.}{2017}]{10.1093/mnras/stx2227}
Stalevski M.,  Asmus D.,   Tristram K. R.~W.,  2017, \mn@doi [Monthly Notices
  of the Royal Astronomical Society] {10.1093/mnras/stx2227}, 472, 3854

\bibitem[\protect\citeauthoryear{Stalevski, Tristram  \& Asmus}{Stalevski
  et~al.}{2019}]{10.1093/mnras/stz220}
Stalevski M.,  Tristram K. R.~W.,   Asmus D.,  2019, \mn@doi [Monthly Notices
  of the Royal Astronomical Society] {10.1093/mnras/stz220}, 484, 3334

\bibitem[\protect\citeauthoryear{{Tristram}, {Burtscher, Leonard}, {Jaffe,
  Walter}, {Meisenheimer, Klaus}, {H\"onig, Sebastian F.}, {Kishimoto, Makoto},
  {Schartmann, Marc}  \& {Weigelt, Gerd}}{{Tristram} et~al.}{2014}]{refId02}
{Tristram} {Burtscher, Leonard} {Jaffe, Walter} {Meisenheimer, Klaus} {H\"onig,
  Sebastian F.} {Kishimoto, Makoto} {Schartmann, Marc}  {Weigelt, Gerd} 2014,
  \mn@doi [A\&A] {10.1051/0004-6361/201322698}, 563, A82

\bibitem[\protect\citeauthoryear{Ueda et~al.,}{Ueda et~al.}{2007}]{Ueda_2007}
Ueda Y.,  et~al., 2007, \mn@doi [The Astrophysical Journal] {10.1086/520576},
  664, L79

\bibitem[\protect\citeauthoryear{{Urry} \& {Padovani}}{{Urry} \&
  {Padovani}}{1995}]{1995PASP..107..803U}
{Urry} C.~M.,  {Padovani} P.,  1995, \mn@doi [\pasp] {10.1086/133630}, \href
  {https://ui.adsabs.harvard.edu/abs/1995PASP..107..803U} {107, 803}

\bibitem[\protect\citeauthoryear{{Uttley}, {Cackett}, {Fabian}, {Kara}  \&
  {Wilkins}}{{Uttley} et~al.}{2014}]{2014A&ARv..22...72U}
{Uttley} P.,  {Cackett} E.~M.,  {Fabian} A.~C.,  {Kara} E.,   {Wilkins} D.~R.,
  2014, \mn@doi [\aapr] {10.1007/s00159-014-0072-0}, \href
  {https://ui.adsabs.harvard.edu/abs/2014A&ARv..22...72U} {22, 72}

\bibitem[\protect\citeauthoryear{{Venanzi}, {H{\"o}nig}  \&
  {Williamson}}{{Venanzi} et~al.}{2020}]{2020ApJ...900..174V}
{Venanzi} M.,  {H{\"o}nig} S.,   {Williamson} D.,  2020, \mn@doi [\apj]
  {10.3847/1538-4357/aba89f}, \href
  {https://ui.adsabs.harvard.edu/abs/2020ApJ...900..174V} {900, 174}

\bibitem[\protect\citeauthoryear{{Wada}}{{Wada}}{2012}]{2012ApJ...758...66W}
{Wada} K.,  2012, \mn@doi [\apj] {10.1088/0004-637X/758/1/66}, \href
  {https://ui.adsabs.harvard.edu/abs/2012ApJ...758...66W} {758, 66}

\bibitem[\protect\citeauthoryear{{XRISM Science Team}}{{XRISM Science
  Team}}{2020}]{xrismscienceteam2020science}
{XRISM Science Team} 2020, arXiv e-prints, \href
  {https://ui.adsabs.harvard.edu/abs/2020arXiv200304962X} {p. arXiv:2003.04962}

\bibitem[\protect\citeauthoryear{Yaqoob \& Murphy}{Yaqoob \&
  Murphy}{2011}]{10.1111/j.1365-2966.2010.17902.x}
Yaqoob T.,  Murphy K.~D.,  2011, \mn@doi [Monthly Notices of the Royal
  Astronomical Society] {10.1111/j.1365-2966.2010.17902.x}, 412, 277

\makeatother
\end{thebibliography}



\section*{Acknowledgements}
JM acknowlages support from NASA grant 80NSSC20K0038. CR acknowledges support from the Fondecyt Iniciacion grant 11190831. We would also like to extend our gratitude to Carolina Andonie for her valuable insights into these simulations. We also thank the referee for their very quick report on this manuscript and many suggestions for improvement.

\section*{Data Availability}
The datasets generated and/or analysed in this study are available from the corresponding author
on reasonable request.

\appendix

\section{Tables for Spectral Line Widths}

\begin{table*}
\caption{\label{table1}Selected spectral line equivalent widths for a torus + hollow cone at $85^{\circ} \leq i \leq 90^{\circ}$.}
\begin{tabular}{lllcccc}
Element & Transition             & Energy (eV)   &  \multicolumn{4}{ c }{EW (eV)} \\
        &                        &       & $N_{\text{H}} = 10^{21}$cm$^{2}$                      & $N_{\text{H}} = 10^{21.5}$cm$^{2}$                        & $N_{\text{H}} = 10^{22}$cm$^{2}$                      & $N_{\text{H}} = 10^{22.5}$cm$^{2}$                        \\ \hline
O       & K$\alpha_{1}$/$\alpha_{2}$                     & 524.9   & 507                 & 602                   & 785                 & 944                     \\
F       & K$\alpha_{1}$/$\alpha_{2}$                    & 625.84  & 44                & 61                  & 81                 & 126                   \\
Ne      & K$\alpha_{1}$/$\alpha_{2}$                     & 848.6   & 163                 & 174                   & 243                 & 321                   \\
Mg      & K$\alpha_{1}$/$\alpha_{2}$                     & 1253.6   & 92                 & 103                    & 125                 & 170                   \\
Si      & $\alpha_{2}$KL$_{II}$  & 1739.38  & 195                 & 165                   & 216                 & 203                   \\
S       & $\alpha_{2}$KL$_{II}$  & 2306.64  & 164                 & 122                   & 41                 & 81                   \\
Ar      & $\alpha_{2}$KL$_{II}$  & 2955.63  & 226                 & 127                   & 47                & 40                  \\
Ca      & $\alpha_{2}$KL$_{III}$ & 3691.68  & 473                 & 253                    & 100                 & 55                 
\end{tabular}
\end{table*}

\begin{table*}
\caption{\label{table3}Selected spectral line equivalent widths for a torus + hollow cone at $45.6^{\circ} \leq i \leq 50.6^{\circ}$.}
\begin{tabular}{lllcccc}
Element & Transition             & Energy (eV)   &  \multicolumn{4}{ c }{EW (eV)} \\
        &                        &       & $N_{\text{H}} = 10^{21}$cm$^{2}$                      & $N_{\text{H}} = 10^{21.5}$cm$^{2}$                        & $N_{\text{H}} = 10^{22}$cm$^{2}$                      & $N_{\text{H}} = 10^{22.5}$cm$^{2}$                        \\ \hline
O       & K$\alpha_{1}$/$\alpha_{2}$                     & 524.9   & 473                 & 482                   & 491                 & 523                     \\
F       & K$\alpha_{1}$/$\alpha_{2}$                    & 625.84  & 49                & 50                   & 49                 & 49                   \\
Ne      & K$\alpha_{1}$/$\alpha_{2}$                     & 848.6   & 179                 & 174                   & 179                 & 179                   \\
Mg      & K$\alpha_{1}$/$\alpha_{2}$                     & 1253.6   & 109                 & 107                    & 109                 & 109                   \\
Si      & $\alpha_{2}$KL$_{II}$  & 1739.38  & 176                 & 172                   & 174                 & 174                   \\
S       & $\alpha_{2}$KL$_{II}$  & 2306.64  & 122                 & 120                    & 109                 & 125                   \\
Ar      & $\alpha_{2}$KL$_{II}$  & 2955.63  & 41                 & 42                   & 41                & 41                  \\
Ca      & $\alpha_{2}$KL$_{III}$ & 3691.68  & 40                 & 39                    & 38                 & 38                 
\end{tabular}
\end{table*}

\begin{table*}
\caption{\label{table4}Selected spectral line equivalent widths for a torus + filled cone at $80^{\circ} \leq i \leq 85^{\circ}$.}
\begin{tabular}{lllcccc}
Element & Transition             & Energy (eV)   &  \multicolumn{4}{ c }{EW (eV)} \\
        &                        &       & $N_{\text{H}} = 10^{21}$cm$^{2}$                      & $N_{\text{H}} = 10^{21.5}$cm$^{2}$                        & $N_{\text{H}} = 10^{22}$cm$^{2}$                      & $N_{\text{H}} = 10^{22.5}$cm$^{2}$                        \\ \hline
O       & K$\alpha_{1}$/$\alpha_{2}$                     & 524.9   & 548                 & 697                   & 864                 & 1181                     \\
F       & K$\alpha_{1}$/$\alpha_{2}$                    & 625.84  & 49                & 71                    & 112                 & 148                    \\
Ne      & K$\alpha_{1}$/$\alpha_{2}$                     & 848.6   & 249                 & 229                    & 294                 & 429                   \\
Mg      & K$\alpha_{1}$/$\alpha_{2}$                     & 1253.6   & 145                 & 107                    & 149                 & 210                   \\
Si      & $\alpha_{2}$KL$_{II}$  & 1739.38  & 260                 & 246                   & 204                 & 243                   \\
S       & $\alpha_{2}$KL$_{II}$  & 2306.64  & 904                 & 311                     & 168                 & 160                   \\
Ar      & $\alpha_{2}$KL$_{II}$  & 2955.63  & 931                 & 312                   & 119                & 61                  \\
Ca      & $\alpha_{2}$KL$_{III}$ & 3691.68  & 1700                 & 917                    & 284                 & 119                 
\end{tabular}
\end{table*}

\begin{table*}
\caption{\label{table6}Selected spectral line equivalent widths for a torus + filled cone at $45.6^{\circ} \leq i \leq 50.6^{\circ}$.}
\begin{tabular}{lllcccc}
Element & Transition             & Energy (eV)   &  \multicolumn{4}{ c }{EW (eV)} \\
        &                        &       & $N_{\text{H}} = 10^{21}$cm$^{2}$                      & $N_{\text{H}} = 10^{21.5}$cm$^{2}$                        & $N_{\text{H}} = 10^{22}$cm$^{2}$                      & $N_{\text{H}} = 10^{22.5}$cm$^{2}$                        \\ \hline
O       & K$\alpha_{1}$/$\alpha_{2}$                     & 524.9   & 479                 & 479                    & 484                  & 486                     \\
F       & K$\alpha_{1}$/$\alpha_{2}$                    & 625.84  & 50                & 50                    & 51                 & 51                    \\
Ne      & K$\alpha_{1}$/$\alpha_{2}$                     & 848.6   & 175                 & 179                     & 180                 & 177                   \\
Mg      & K$\alpha_{1}$/$\alpha_{2}$                     & 1253.6   & 109                 & 107                    & 109                 & 109                   \\
Si      & $\alpha_{2}$KL$_{II}$  & 1739.38  & 174                 & 174                       & 174                              & 176                    \\
S       & $\alpha_{2}$KL$_{II}$  & 2306.64  & 123                 & 123                  & 124                 & 123                    \\
Ar      & $\alpha_{2}$KL$_{II}$  & 2955.63  & 41                 & 41                   & 41                & 41                  \\
Ca      & $\alpha_{2}$KL$_{III}$ & 3691.68  & 37                 & 38                    & 39                 & 39                 
\end{tabular}
\end{table*}

\begin{table*}
\caption{\label{table7}Selected spectral line equivalent widths for a torus + hollow cone at $85^{\circ} \leq i \leq 90^{\circ}$ with an opening angle of $\alpha = 10^{\circ}$.}
\begin{tabular}{lllcccc}
Element & Transition             & Energy (eV)   &  \multicolumn{4}{ c }{EW (eV)} \\
        &                        &       & $N_{H} = 10^{21}$cm$^{2}$                      & $N_{H} = 10^{21.5}$cm$^{2}$                        & $N_{H} = 10^{22}$cm$^{2}$                      & $N_{H} = 10^{22.5}$cm$^{2}$                        \\ \hline
O       & K$\alpha_{1}$/$\alpha_{2}$                     & 524.9   & 432                 & 471                   & 568                 & 717                     \\
F       & K$\alpha_{1}$/$\alpha_{2}$                    & 625.84  & 41                & 95                  & 121                 & 162                   \\
Ne      & K$\alpha_{1}$/$\alpha_{2}$                     & 848.6   & 143                 & 169                   & 234                 & 331                   \\
Mg      & K$\alpha_{1}$/$\alpha_{2}$                     & 1253.6   & 122                 & 130                    & 143                 & 226                   \\
Si      & $\alpha_{2}$KL$_{II}$  & 1739.38  & 140                 & 196                   & 216                 & 275                   \\
S       & $\alpha_{2}$KL$_{II}$  & 2306.64  & 68                 & 250                   & 192                 & 122                   \\
Ar      & $\alpha_{2}$KL$_{II}$  & 2955.63  & 337                 & 183                   & 63                & 42                  \\
Ca      & $\alpha_{2}$KL$_{III}$ & 3691.68  & 1392                 & 429                    & 194                 & 80                 
\end{tabular}
\end{table*}

\bsp	
\label{lastpage}
\end{document}